# Plasma fireball sheath dynamics: A brief review and meta-analysis

Subham Dutta[1], Ahmed Atteya[2], Pralay Kumar Karmakar[1*]
[1]Department of Physics, Tezpur University, Napaam, Tezpur, Assam-784028, India.
[2]Department of Physics, Faculty of Science, Alexandria University, Alexandria 21511, Egypt.
*Email: pkk@tezu.ernet.in

**Abstract**
We present a comprehensive overview of the formation mechanism of plasma fireball sheath (PFS) structures, sheath-induced collective phenomena, associated relevant instabilities, and corresponding onset conditions. It includes an optimum set of self-illustrative schematic figures, relevantly manifesting the instability triggering dynamics, various involved metastable stages, and parametric threshold conditions. The possible damping mechanisms of excited instabilities in the usual PFS systems are also highlighted. An up-to-date experimental glimpse of both the regular fireball (RFB) and inverted fireball (IFB) classes is specifically presented. We explicitly offer illustrative appendices showing the main distinctions between (a) RFB and IFB, (b) laboratory and astrocosmic fireballs, and (c) RFB sheath and IFB sheath formations. It provides a panoptic glimpse of the current RFB and IFB studies with a special mention to both laboratory and astrospace plasmas. A holistic outline on the chronological development of the PFS research investigations, since the inception of plasma-electrode coupling studies, is outlined. A clear indication of the future fireball scope in both theoretic and applied perspectives is finally emphasized.

**Keywords** Plasma fireball. Sheath. Instability. Fireball history

## 1 Introduction

The terminology "Fireball (FB)" represents a generic expression to describe a variety of optical glows and associated excitation-ionization phenomena invariantly prominent in various plasmic and non-plasmic conditions on diversified scales of space and time. Its naturalistic existence spans extensively from the laboratory scale environments [1, 2] to the astrocosmic ones [2, 3] in various realistic circumstances [1-3]. It is seen accordingly that, despite the variation of the formation mechanism irrespective of spatial dimensions, the usual FB model scale-invariantly holds good in various plasma environments. For an instant example, in a typical laboratory plasma chamber, an embedded electrode with its electrostatic potential (due to external biasing) almost raised up to the ionization potential of the background atoms with neutral gas pressure $\sim 10^{-2}$ to $10^2$ mTorr, thermal energy $\sim 0.5$ eV to $2$ eV, electrode potential $\sim 30$ V to $70$ V (approximately) can trigger the formation of a plasma FB around the said electrode [4]. It is noteworthy that cathodes (against anodes) can also be used to generate FBs in such circumstances. But cathodes are rarely used here due to the fact that it is more convenient to mobilize the electrons (by anodes) than the ions (by cathodes) in the electrode periphery for plasma FB creation and subsequent diagnostics (since $m_e/m_i = 5.45 \times 10^{-4}$).

With the variation of the electrode biasing potential and neutral gas pressure, the circumventing sheath changes in terms of its spatial potential variations. These various types of spatial variations are individually named, such as ion sheath, electron sheath, double sheath, anode glow, FB, and so forth. The said FB is an abrupt transient stage which occurs in between the ion sheath and anode glow region. The FB formation is usually observed for a timescale of 10 $\mu s$ and pressure of magnitude 1 mTorr in argon plasma [4, 5]. To add further, the ratio of the plasma chamber wall area ($A_W$) and the electrode area ($A_E$) also influences the transition from the ion to the electron rich sheath, thus also in the formation of the subsequent plasma FB [5]. In this plasma FB context, an up-to-date experimental glimpse of both the regular fireball (RFB) and inverted fireball (IFB) classes is systematically presented later in this manuscript.

It may be noteworthy to mention here that plasma fireball sheath (PFS) formed in a laboratory plasma chamber is dependent upon the neutral gas pressure ($p$), electrode voltage ($\phi$), areal aspect ratio ($A_E/A_W$), choice of the host plasma, etc. Both $p$ and $\phi$ act in favor of developing the FB formation. The relevant areal ratio makes the shape of the electrode very pertinent for creating unidirectional electron-ion flow and for inducing more efficient electron-neutral collisions. For the unidirectional flow, plane electrode geometry is most efficient as done by Baalrud et al. [5]. Although the PFS is modelled to be spherical in shape for its simplified geometry, the planar anode geometry is also found to be efficient in structuring plasma FBs in many experimental situations.

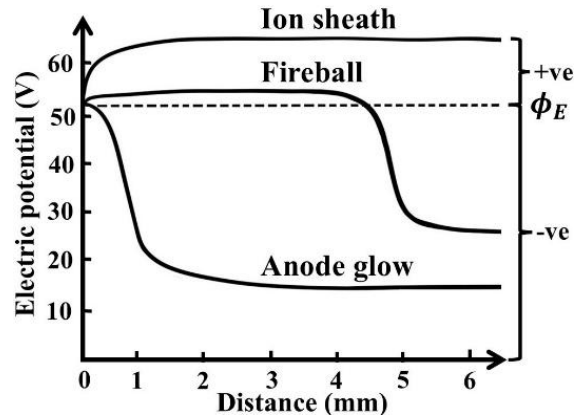

**Fig. 1** Schematic showing electrostatic sheath potential ($\phi$) variation across various kinds of sheaths with respect to distance ($r$) from the electrode ($E$, inspired by Ref. [5]). The $+$ve and $-$ve signs denote positive and negative space charges, respectively.

The variations of the sheath structures occur due to the geometry of the electrode and its biasing. Increasing the bias of the electrode increases the electron kinetic energy (KE)$\sim e(\phi_E - \phi_p)$, where $\phi_p$ is the plasma potential. A charge has KE equivalent to the difference between the electrode potential and the plasma potential. Once the biasing reaches a critical magnitude equivalent (comparable) to the ionization potential of the neutral gas, the ionization cross-section also increases. The ionization yielded electrons and ions spend unequal amounts of time inside the sheath due to their mass difference. The electrons get absorbed by the electrode much earlier than the ions are dispersed outside the sheath due to its opposite electrical polarity. If the ionization rate is high enough (greater than recombination), the number of ions in the sheath supersedes the number of electrons in it. A potential transient region formed from the negative space charge to the positive space charge in the outer periphery of the sheath is termed as a double layer (DL). The ionization and excitation within the sheath produce a dim glow, termed as anode glow, lying between the anode and the DL [5]. It is pertinent to add further that the electron sheath flattens when the anode glow forms and so forth.

Increasing the electrode biasing further, the anode glow bifurcates the sheath structure to a much larger luminous secondary discharge (known as the plasma FB). The DL potential drop is approximately the ionization potential of neutral gas. The DL has a width of the order of the Debye length. The diameter of the FB and required biasing for FB formation is inversely proportional to the neutral gas pressure [5]. Reducing the electrode biasing does not form the anode glow back at the same biasing voltage but at higher voltages. While increasing the biasing, the FB forms for approximately 1 μs at a neutral gas pressure of 1 mTorr in argon plasma [5]. Thus, it may be inferred therefrom that the FB is a special transient region between an anode glow and the ion sheath. For further reading on the transient FB nature, one may refer to the comprehensive review article authored by Baalrud et al. [5].

The formation of the laboratory plasma FBs (spherical glows with diameter~5-10 cm) is attributed to the converging drift of the background electrons via the plasma sheath [4]. This electronic drift results in continuous collisions among the electrons (> 10 eV) and neutrals (< 1 eV) [4]. Consequently, these collisional effects trigger glows originating from the release of collision-induced excitation energy, as illustrated schematically in Figs. 2-3, with the neutrals getting deexcited back to the ground state [6]. Besides, such collisional processes can also lead to secondary ionization processes and ion-acoustic instabilities instantaneously [4]. The ionization also helps the creation of a quasineutral plasma FB as the produced ions cannot be immediately dispersed by the plane anode, and they gather around the anode creating a local maximum therein. The anode-surrounding electrons find the local maximum as a potential well, jumping more into it forming the quasineutral zone of the plasma FB [7]. The PFS is usually assumed to form around a spherical anode; therefore, the spatial variation of the local maximum properties may vary in this case, as it is noticed experimentally that the $A_E/A_W$-value influences the collisional cross section and subsequent outcomes in terms of various relevant physical FB parameters [5].

The glow formation of the FBs can be physically attributed to the inelastic electron-neutral collisions, which, in fact, excite (also, ionize) the latter up to the energy corresponding to that of visible spectra (of excited atoms). The kinetic energy of the colliding electrons originates from the electrostatic potential energy localized across the sheath formed around the employed central anode in the plasma system in accordance with the universal law of conservation of energy. The plasma FB glows because of the release of that collision-induced excitation energy of the neutrals [6, 8].

It may be noteworthy here that, on increasing the electrode potential, there occurs a successive widening of the sheath across the applied electrode, and then, the FB develops within it. With further accumulation of charges in the sheath, the DL forms adjacent to it, but radially outwards. The DL circumvents the sheath as a nonlinear space charge layer with a bipolar structure having an inter-layer width on the Debye-scale order. The DL plays the role as a primary agent to drive outer electrons to the sheath. As a result, it dynamically broadens the sheath width (also, the FB) with coagulation of charges from the ambient plasma and so forth [9].

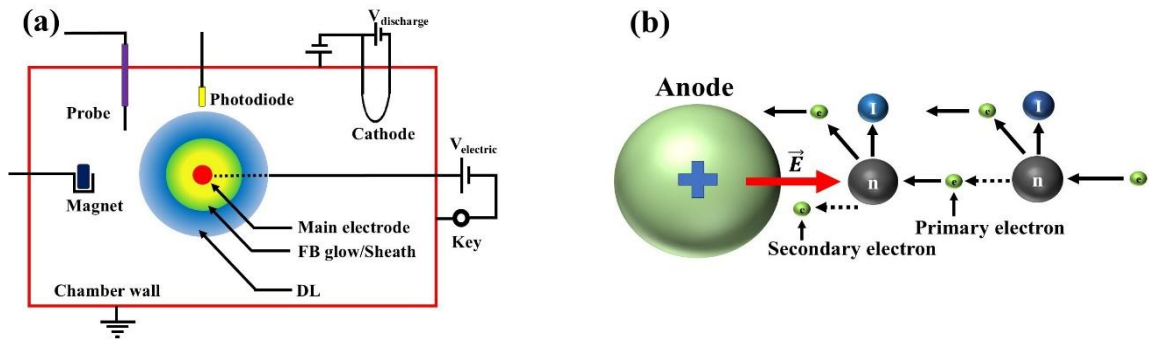

**Fig. 2** Schematic diagram showing (a) plasma discharge setup forming FB and (b) physical mechanism triggering the plasma FB formation (inspired by Refs. [4, 10]).

The volumetric dimension of the plasma sheath is determined by the balancing width of bi-directional electron-ion flux (floating condition) across it during the FB operation. Any imbalance in the interparticle flux leads to the excitation of various instabilities, such as the sheath plasma instability (SPI) [11], two-stream instability (TSI), ionization instability [4], ion-acoustic instability (IAI) [4], and so forth. Such instabilities manifest a wide spectrum of utilities in fundamental plasma physics, material sciences, etc. [1]. These instabilities are influenced by the DL formed around the sheath. The prevalent electrostatic potential difference across the DL during the FB glow formation is equivalent to the ionization potential of the host gas atoms [9, 12]. It is obvious that the FB glow does not develop with any electrode potential lower than that of the required threshold ionization energy value of the neutral atoms.

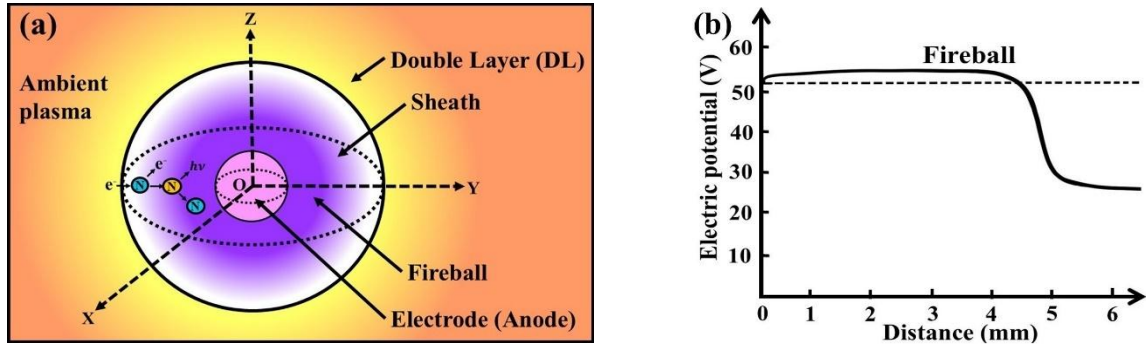

**Fig. 3** Schematic diagram artistically displaying (a) usual PFS structure and (b) spatial electric potential variation across it. The central sphere (light pink) in the former is the anode. The circumvent sphere (violet, fading outwards) is the sheath forming the RFB.

In Fig. 3(a), the color gradually fading outwards across the sheath signifies the variation in the electric charge polarity in the sheath interior (inner) region. The exterior (outer) region (yellow) is the corresponding DL structure. The background reddish zone denotes the ambient plasma in the adopted experimental setup.

In the current context of FB structurization, a special FB formation, unlike all those already discussed above, within a reticular plasma submerged anode has recently been reported [13, 14]. This special category of the enveloped (by anode) FBs is classified as an IFB, mentioned earlier [8]. The fundamental difference between RFBs and IFBs is that the earlier envelops the anode volumetrically (without an applied magnetic field); whereas the latter is enveloped by the anode. The similar adjacent regions in RFBs, viz., sheath and DL, also form across the enveloping IFB anode on its outer side (Fig. 4(a)). The sheath thickness is determined by the anode grid width. Although the IFB formation dynamics are same as that of the RFB; but, the IFBs deal with varieties of excited atypical collective waves, eigenmodes, and instabilities along with their widespread implications and applications [1]. A simple comparison distinguishing the two distinct RFB and IFB is presented in a tabular form in Table-1. The basic distinctions between the laboratory FBs and astrocosmic FBs are concisely tabulated in Table-2. Then, the regular plasma sheath is contrasted with the inverted plasma sheath structures in Table-3.

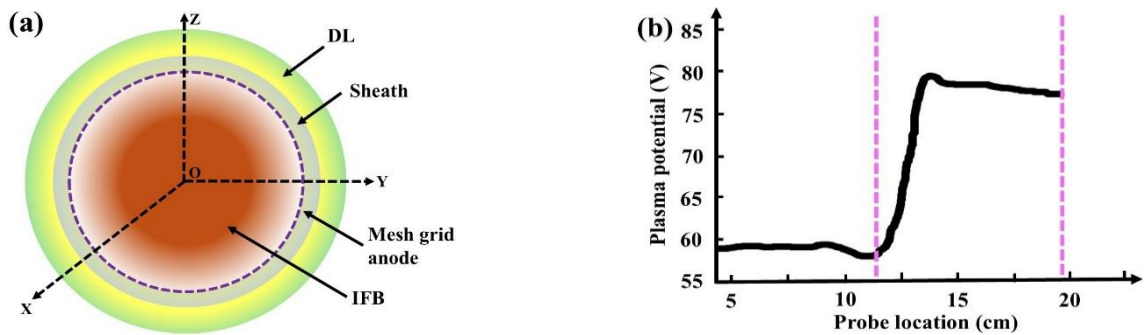

**Fig. 4** Schematic diagram showing the (a) formation of an IFB within a reticular electrode (anode) and the corresponding (b) electrostatic potential profile (inspired by Ref. [8]).

In Fig. 4(a), the various distinguished layers denote (inner to outer) the IFB glow (orange gradient), anode (violet dashed circle), IFB sheath (light grey), DL (yellow-green gradient). In Fig. 4(b), the pink lines denote the IFB boundary. Besides, both electron and ion temperatures are constant within the equipotential IFB structure [8].

It is noteworthy that FB modelling is limited not only to laboratories and above mentioned astroplasmic scenarios, but it is also relevant in several other plasmic domains of both pure and applied values. Some of those widespread plasmic fields of research, based on FB models, are briefly discussed herein. For example, the FB model helps in understanding the meteor radio afterglows (MRAs) in the meteor trails [15]. These meteors also form bright glows while drifting across the sky. These glowing meteors can be modelled as astrocosmic FB, since they possess a brightness of magnitude $-3$ to $-5$ for a corresponding meteoroid of mass $10 \text{ g}$ only [16]. The MRAs originate from meteoritic movements, and some come from diffuse galactic regions also. They may be observed after reflection from the mirroring meteor trails through various observation points [15]. These meteoritic FBs, however, are not known to possess any electrostatic sheath or DL as observed in the laboratory FBs. Besides, modelling such meteoritic events as FBs helps in exploring their multiple mechanical properties, such as their fragmentation (e.g., Geminid meteor shower), composition, mass, brightness, and other relevant physical properties [16].

In the relativistic regime, the FB model is utilized in studying the production of anti-nucleons through the standard model (SM) particle physics, studied commonly in high-energy physics [17]. Any abrupt localized injection of large amounts of energy and anti-baryonic matter in the form of SM anti-quarks can trigger the formation of locally thermalized FBs. These FBs comprise of plasma mixture of free anti-nucleons, pions, leptons, and emitted photons [17]. With the hydrodynamic expansion of the FB, they cool adiabatically allowing anti-nucleosynthesis to generate bound anti-nuclei, including anti-Helium in the same hot and denser surroundings [17].

To add further, in an explosive direction, the nuclear explosion (atom bomb) has also been fruitfully modelled as a FB structure, highly helpful in understanding the warhead science to encounter and overcome any probable nuclear threats with minimum damage [18].

Along with the non-plasmic FB models, it may also be noted that substantial progress has been made in the study of laboratory PFSs, associated stability phenomena, and their consequent applications extensively [1]. However, it specifically includes complex PFS-centric scenarios in the diversified astrolabcosmic contexts [2].

It may be highlighted that the most relevant experimental research on plasma FB has been conducted during the last decade by Prof. Stenzel et al. in DC discharge plasmas [1, 2, 4]. Although the number of research works in the plasma FB direction has grown less rapidly in the literature in the current decade, the technological pertinence of the FB research area has grown more rapidly [1]. Several technological applications of the FB research, such as, in electronics, jet propulsion, astrophysical modelling, etc. [1-10] have been elaborated afterwards to prove its importance. Besides, the pertinence of FB research across the fundamental and applied fields is the main motivation for this review work with a special mention to the future scope.

A chronological report-based thematic overview seems to play a significant role in the developmental process of conceptional foundation about the PFS and IFB research scenarios so far. This contribution presents an integrated overview on the active PFS dynamics in a broader systematic horizon bearing technological relevancy vertically. It chronologically presents the noteworthy advancements in this PFS and IFB fields of research alongside their significant outcomes reported so far. Lastly, the paramount implications and tentative applications of our study together with non-trivial future scope for further investigation in this research direction are outlined summarily.

## 2  PFS research growth

In a historical perspective, it is well-known that plasma FB phenomena founded on the collective interactions between the plasma and submerged electrode, have been reported first by O. Lehman in 1902 [19]. Such interaction phenomena have been further explored broadly by I. Langmuir, H. M. Mott-Smith, and L. Tonks [20, 21]. Thus, it is pertinent to discuss briefly the beginning of plasma electrode interaction mechanism, which began soon after the discovery of plasma in laboratory. Although it was discovered in 1879 by W. Crooks, the term, 'plasma', has been coined later in 1929 by Langmuir for some specific ionized gases fulfilling certain criteria, as already well known from the literature [22]. It is seen that the previous plasma-electrode analyses (which lead to the discovery of FB glow subsequently) have broadly accounted for the collector current variation with the applied electrode voltage [20]. Such investigations have been conducted in low-pressure ionized gas chambers with the corresponding electric current collected by a plasma-inserted anode, biased externally.

The PFS study gradually has initiated comprehensive reporting on the sheath formed around the inserted electrode. Sheath physics has been studied in terms of the spatial variation of electrostatic potential, velocity (of charged particles), space charge density, sheath width, sheath potential, collector current through it, and so forth. If the electrode potential is higher than that of the ionization potential of the host gas, the sheath can be classified into two parts in terms of the rate of spatial potential variation. The central part manifests the maximum potential variation comprising of opposite charges (against the electrode potential) and few same polarity charges of adequately high thermal velocity (to the electrode potential) in the outer region. Although, there is no sharp edge to demarcate the sheath regions sharply and strictly, there is an outer diffused edge of the steep potential which can be considered as the outer sheath boundary. This outer part comprises of bipolar electric charges with a feeble potential variation. The formation of the PFS is consequently a special case of the plasma-electrode interaction mechanism. The FB glow may not form at a lower electrode potential, but the sheath around the plasma submerged electrode is bound to form as an intrinsic plasma property [23, 24].

Apart from the spatial potential variations, the interrelation of the dimensional variation of plasma sheath with the help of the Langmuir probe has also been investigated by L. J. Sonmor and J. G. Laframboise in 1991. The plasma-electrode system analyses in magnetic field have been introduced in 1949 by D. Bohm et al. [26]. The sheath structure was found to deviate from its usual shape in the presence of magnetic field, as the fast-moving electrons and ions across the sheath interact with it [25]. The advancing literature published ahead (during 1960-70) have encompassed the outcomes of the studies on the spatiotemporal fluctuations of the current, voltage, and various other related plasma parameters in both magnetized and unmagnetized plasma environments [26]. A few of such results conducted by Stenzel et al. have been presented in Fig. 5, which shows the temporal variation of the electric current, voltage, and light emission.

The widespread research investigations on the induced electron-ion cyclotron and acoustic oscillations in the plasma-electrode system have commenced in 1963 with the results of R. W. Motley and N. D'Angelo [27]. Such acoustic oscillations are also pictorially illustrated in Fig. 11 manifesting their spatial variation. A vast area of various nonlinear eigenmodes and wave formations in the plasma-electrode system is yet to be explored both theoretically and experimentally. With successive progress, predominantly in the theoretical sector, the PFS has also been analyzed through both linear and nonlinear formalism. Analyses on various instabilities pertaining to the plasma sheath have been initiated by R. L. Stenzel in 1989 [28]. Such studies on plasma sheath have begun with the introduction of a temporally fluctuating or pulsed electrode in 1963 by D. G. Bills et al. [29]. Similar studies have been continued ahead by J. M. Urrutia and R. L. Stenzel [30], and others. The current flow to the pulsed FB from the ambient plasma has been witnessed to occur through the whistler and Alfven-like nonlinear electromagnetic eigenmodes (as well-illustrated in the section of 'Linear normal PFS wave modes') [30, 31]. These eigenmodes have been found to originate due to the plasma electron-ion density fluctuations around the RFBs and IFBs. The excited eigenmodes then develop as DLs around the PFS [1]. The DL, due to its intrinsic electric field, accretes the adjacent charge particles towards the sheath. The associated physics behind PFS is invariantly relevant over diversified spatiotemporal scales ranging extensively from the nanoscopic to cosmic domains, and so forth [1-3].

It has taken a couple of decades to develop an experimental understanding about the various instabilities originating across the sheath, around a plasma submerged anode, specifically around a PFS. As far as seen in the literature [13, 14, 32, 33], the existence of PFSs has been reported first in 2011 [13, 14], followed by detailed analyses (by R. L. Stenzel) on plasma FB instability, including its transit time evolution [13, 14]. Besides, several other plasma instabilities are found to get excited across astrophysical domains, such as ionization instability, ion-acoustic and dust-ion-acoustic instabilities [34], two-stream instability [35, 36], Kelvin-Helmholtz instability [37], Rayleigh-Taylor instability [38], and so forth. A concise review of the various instabilities operative in various naturalistic plasma systems is being presented later illustratively.

Apart from the RFB description above, it is seen historically that the first reporting of the IFB structure can be traced back to 2010 (the same year as the RFB) by R. L. Stenzel et al. [13]. Some dominant vircator-like instabilities in its formation and high-frequency oscillations with typical time periods (~ electron transit time) through the reticular spherical anode have experimentally been reported [13, 14]. These oscillations do not indulge the plasma eigenmodes owing to their lower frequencies in comparison to the electron plasma frequency. These oscillations fall weaker to match with that of the threshold magnitude of natural eigenmode frequencies and hence, finally, get damped due to the dissipative effects pertaining to multiple damping agents present therein [13].

The entire PFS physics has been comprehensively studied by different researchers in diversified plasma environments with various laboratory and experimental setups. A brief chronological highlight of various studies on the PFS system and their respective outcomes in different configurations can be concisely summarized as follows.

## 2.1 PFS with different electrodes

### 2.1.1 PFS with constant voltage electrode

The simplest and more general way of producing a PFS is through having constant electrode voltage ($\sim 50 - 100$ V) with adequate pressure ($\sim$ mTorr) and number density ($10^{14} - 10^{17}$ m$^{-3}$). The FB can be generated by gradually increasing the electrode voltage with the other two parameters kept constant for a specific gas. These typical values of the parameters vary with the variation in the gas type [4]. The light emission from the FB is examined with a photodiode mounted outside the plasma chamber by the current reading through it.

It is interesting that even with constant voltage biasing of the FB centered electrode the current collected through the FB by the probe shows unstable behavior with periodic peaks. The repetition time of the peaks are of $\sim 100$ μs. The reason for these peaks may be explained by the generation of primary and secondary ions through ionization due to electron-neutral collisions across the sheath. The electrons get absorbed by the electrode whereas the ions experience a diverging push. Due to the lower mobility of the ions, they are not immediately dispersed out but are observed to form small bunches to be eventually released from sheath forming the repetitive current pulses [4].

### 2.1.2 PFS with pulsed electrode

One of the ingenious ways of analyzing the PFS glow (both in RFB and IFB), associated fluctuations, waves and instabilities is through introducing a pulsing central electrode (anode forming the FB) in the experimental plasma system. The pulsed FB glow and the corresponding collected current have been widely discussed in the literature [4, 6, 9].

The instantaneous current disruptions due to pulsation generate the electron and ion cyclotron waves along with the formation of a DL [4]. The DL further coagulates more electrons (ions) in the FB periphery. The DL gradually widens with the reflected electrons (ions) from the sheath [9]. The observed FB glows and emanated current pulses are attributable to the pulsed background plasma production in the presence of the collisional ionization of the neutrals (Figs. 2-3). It results in a pulsed supply of electrons colliding repeatedly with the neutrals [9]. The collector current collection (through a probe) with the temporally pulsed electrode potential is found to generate background plasma with every successive pulse. The successive pulse rates subsequently enable us to determine various quintessential plasma parameters, such as the plasma potential, electron temperature, density, and so forth [39]. The electric currents and FB glow fluctuations are the outcomes of discontinuous plasma production with temporally varying electron supply resulting due to the pulsating central anode potential [8].

One of the advantages of producing a pulsed FB at sufficiently high pulse rate is that with every successive FB formed, the afterglow plasma produced is sufficient in igniting the next FB. This pulsed arrangement, hence, nullifies the need of a background plasma (BGP) source. The pulse rate should be high enough to prevent the ionized elements neutralizing through recombination. Which otherwise would need further ionization for BGP production [4]. In other words, no BGP source is needed having a higher frequency pulse rate of the electrode voltage.

In addition, one of the special outcomes of pulsed electrode diagnosis of PFS is an induced anharmonicity between the electric field pulses and corresponding electron drifts. This anharmonicity is a yield of the weak but finite nonzero electron inertia, which significantly attenuates the sheath oscillations as an intrinsic resistive property. Electron inertia is also responsible for the sheath plasma and relaxation instabilities in the presence of constitutive drifting ions [11, 40]. If both the field and the electrons manage to oscillate in a synchronized rhythm, the electrons positively feedback the electric field fluctuation across the sheath. This specific inertia-triggered instability is called "Inertial instability" or "Transit time instability" in the literature [41, 13, 14]. A special type of such relaxation instabilities due to a pulsing probe is the potential relaxation instability (PRI), arising because of weak but finite electron inertia [40]. This instability occurs due to the rhythmic charge density variations around an electrode [40]. Both the anode electric field and KE of the charges drifted by it fluctuate spatiotemporally.

The direction of electric field across a well-grown plasma sheath has been reported to reverse (phase overturn by 180°), i.e., to lie in an anti-phase configuration [42]. This directional overturn in the electric field under some specific circumstances is referred to as the "plasma sheath inversion". A few of the accountable circumstances for this directional overturn of the field comprise new particle inclusion, pressure, and density fluctuations around the sheath, and so forth [43, 44]. An analytic investigation manifesting the changes in the potential energy (PE) and KE flux of the electrons and ions in the inverse sheath has been reported recently [42-44]. In this context, a table showing the main distinctions between the regular and inverted sheaths has been additionally presented in Table-3.

The fluctuation in light emission in Fig. 5 (red curve) occurs due to multiple de-excitations of the excited electrons from the lower excited states of the host plasma atoms. The light saturation at higher voltage may account for higher degrees of ionization, resulting in higher luminosity of the emitted light, but less frequent and uninterrupted eventual de-excitation.

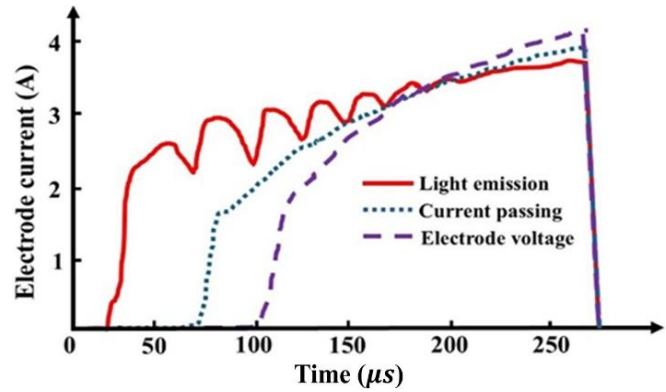

**Fig. 5** Manifestation of temporal variation of electrode current during formation of a pulsed FB within a single pulse. Variations of electrode voltage, current collection, and light emissions are highlighted in violet, blue, and red color curves, respectively. The voltage along the vertical axis ranges from $\sim 30 - 80$ V (inspired by Refs. [4, 5]).

### 2.1.3 Comparison of PFS with different electrodes
The PFS is stable without any instability and for constant plasma parametric values (such as electrode biasing, pressure, density, etc.) up to certain range. Beyond which sheath plasma oscillations start to occur which may take the form of SPIs. Moreover, drawing currents from plasma also produces electron and ion acoustic waves in the medium. Thus, the stability of plasma is a special case, whereas the dynamic nature is an almost general case.

    Thus, the criteria for a stable PFS can be experimentally found out in terms of critical magnitudes of potential biasing, pressure, magnetic field strength (if applied), areal ratio, etc. The deviation from the required range of magnitudes of these parameters yields a dynamic PFS.

    The PFS described in the Baalrud et al. (Ref. [5]) work is generated using a solid anode with special plane anode geometry. Thus, in terms of the anode morphology and geometry the discussed PFS is a special class of it and can vary from this special arrangement when the inverted fireball (IFB) geometries with varying electrode biasing (within the electrode itself) are considered. Since spatial potential variation across the two classes of FBs are different. The entire glow region in RFB has a spatially varying potential, whereas, in IFBs, the center region of it possesses a spatially invariant potential domain, quite resembling a Faraday cage.

### 2.2 PFS with diagnostic tools
It is noteworthy that one of the most convenient ways of diagnosing the PFS system is through introducing a plasma probe (e.g., Langmuir probe, double probe, etc.) across it. The current-voltage characteristics measured through the probe reveal multiparametric information about the PFS system, e.g., the electron-ion density, plasma temperature, electrostatic potential, particle mobility, etc. [8]. It may be worth mentioning here that a pulsed probe may reliably be used for an accurate measurement of the electron temperature and electron population density [41]. The size, shape, and location of the FB is influenced significantly by the probe perturbation founded on the plasma-wall interaction processes. It may disturb plasma production and influence plasma loss (due to electron-ion recombination processes) as well [6]. An unstable FB outbursts bunches of constitutive charged particles and associated collective waves. This sudden outburst instantly destabilizes the FB surroundings [4]. The instantaneous spatiotemporal fluctuations in the electron-ion densities inflict an imbalance between the plasma production and loss processes. The electron-ion density fluctuations across the FB lead to the development of relaxation oscillations during its operation and diagnosis [6].

The diagnosis of astrocosmic FBs is, on the contrary, done through the astro-spectrometric observation methods. Since the meteors forming the FBs burn out in the Earth's atmosphere due to the increasing air density downwards, with rare traces of their remnants reaching at the Earth's surface for laboratory experimentations [45].

## 2.3 Nonlinear coherent structures

A wide coagulation of charges, opposite in polarity against the applied electrode biasing around the central electrode in a plasma system, forms the plasma sheath (width ~ 5-10 mm). As already well-known, the plasma sheath is a nonlinear coherent structure in nature, which may, indeed, induce sheath-plasma instability in the PFS system [1, 8]. The formation of a plasma sheath around the electrode (both in RFB and IFB) further induces another non-neutral zone, designated as DL. In this context, it may be mentioned that sheath is a direct outcome of the lowest-order nonlinearity, whereas the DL is a consequence of the higher-order nonlinearities in any system.

It is pertinent to discuss here the formation mechanism of a DL. As a result of ionization across the sheath, more background plasma is created in the form of fresh electrons and ions. The electrons immediately approach the anode for absorption whereas the ions experience an outward push. However, due to the ions being comparatively immobile, they lag in dispersing outwards, eventually forming a higher potential region just outside the sheath. This special zone with sudden variation of electric potential across the junction of electron-rich sheath and accumulated ions is termed as DL. This DL is highly nonlinear in terms of spatial potential variation. It also helps in accelerating the electrons through it to collide with the neutrals with sufficiently high velocity to ionize them and repeat the ionization process over. The potential variation across the DL is of the order of the ionization potential of the host plasma. Therefore, as an example, this DL potential is $\sim 15\ eV$ for Argon plasma and $\sim 24\ eV$ for Neon plasma [4]. The DL is also referred to as 'anode glow' and its potential variation is shown in Fig. 1 in reference [5].

The drifting charges (fractionally) of opposite polarity towards the enclosed sheath collide inelastically with the neutrals in the path [4]. Some of the primary electron-neutral collisions result in the secondary ionization of the neutrals liberating fresh ions and electrons as new plasma components. The liberated charges of opposite polarity join the sheath in no time due to the electrostatic attraction. The addition of electric charges widens the sheath thickness, and it, consequently, gets bulkier, eventually developing an adjacent DL outward [4]. The DL bifurcates the FB from the ambient plasmas. We see that the DLs also excite several instabilities having practical importance [1]. It includes relaxation instabilities, electron transit time instabilities, and so forth. It is noteworthy that the development of both the DL-induced instabilities and widening FB is attributable to the secondary ionization and the background plasma production due to the incessant electron-neutral collisions in the PFS systems [1].

It may be noted that the IFB (Fig. 4) forms within the grid anode and is internally field-free unlike the RFBs (Fig. 3). As a result, no plasma sheath or DL develops inside the anode of the IFB under study. Therefore, we can summarily infer that the different FB arrangements in the two cases (RFB and IFB) yield different and asymmetric potential structures with individual practical relevance in their respective extensive fields of interdisciplinary relevance [4, 8].

## 2.4 Magnetic field effects in PFS dynamics

Application of an external magnetic field in a plasma system excites a plethora of collective waves and relaxation instabilities of practical importance ubiquitously encountered in diversified spheres, such as the PFS dynamics of current concern, planetary magnetosphere, and so forth [1, 45]. A deformation (radial) and rotation (polar and azimuthal) of the PFS are observed due to the resultant engulfing composite ($\vec{E} \times \vec{B}$) field across the PFS ($\vec{E}$: electric field, $\vec{B}$: magnetic field). The magnetic field shields and diverges the electrode-oriented current. The collector current reduces significantly due to diversion of the constituent particles [25]. The relaxation instabilities and the geometrical fluctuations of the PFS dimensions so caused can be diagnosed with the help of the temporally varying magnetic field [9].

More recently, in the collisionless DC discharge magnetized plasmas, a new mode of oscillation, termed as the Mixed Mode Oscillation (MMO), has been reported to originate due to certain synchrony of applied grid voltage and magnetic field magnitudes across the IFBs. The MMOs are complex collective disturbances especially comprising of aperiodic number of small-amplitude (slow) and large-amplitude (fast) oscillatory motions. It is relevant in this context that the consecutive fluctuations of the applied magnetic field and the probe positions typify the MMO yield across the IFB system [47, 48]. In addition, the Alfven mode, whistler mode, and other magnetohydrodynamic (MHD) modes are also found to originate in the PFS systems under test [1, 4]. In such circumstances, it is admitted that both whistler and Alfven modes may play as the normal modes yet to be dealt well in almost realistic MHD environments elaborately. The formation mechanism of such diverse modes is briefly highlighted in the next section. A systematic analysis of such modes in a laboratory plasma based PFS systems can assist further progress in astrophysical science and epitaxial manufacturing technology.

## 2.5 Linear normal PFS wave modes

The linear normal wave modes dominant in a typical magnetized PFS system comprise mainly of the electron-ion-acoustic and cyclotron waves [26, 27]. It is noteworthy that the normally dominant distinct eigenmodes in magnetized FB configurations are the whistler and Alfven waves [30, 31]. These eigenmodes can also carry current across the FBs [4]. A stream of fast electron and ion beams injected into the magnetized PFS through electromagnetic interaction with the pervading magnetic field may yield a plethora of acoustic waves in the PFS system as well [49].

Drawing electric current from the plasma chamber can also lead to plasma oscillations near the ion cyclotron frequency. The electric field across the DL drifts the electrons in the plasma with respect to the background ions. The plasma may also undergo enough disturbances to form ion cyclotron wave in the presence of externally applied magnetic field [27].

The induced electric field across the DL intends to sway both the electrons and ions alongside. However, the agile electrons drifted more instantly than the heavier inertial ions due to their greater mobility. This anharmonicity in the drift is responsible for the under-damped FB oscillations. The oscillatory dispersive AC current due to the inharmonious electron-ion movement precedes the slower DC current triggering the whistler mode in the plasma [30].

The whistler waves are electromagnetic waves (frequency ~ $3 \times 10^2$-$3 \times 10^4$ Hz) generated by electron driven instabilities (e.g., PFS system). They travel along the direction of the magnetic field with luminal velocity. In space plasmas, particularly around the Earth, the whistler waves propagate through ionospheric and magnetospheric region making a whistling sound. The whistling sound originates because of the reflected high-frequency wave arriving at the amplifier before the low-pitched signals. Whistler tones are found to vary in a wide-range spectrum moving from dispersive laboratory to space plasma regions [50].

In addition to the above, another very low-frequency magnetohydrodynamic (MHD) wave, termed as the "Alfven wave", may occur in the magnetized plasmas along with an FB formation. The Alfven wave originates due to the perturbation of magnetic field and velocity of the electron-ion currents perpendicular to the magnetic field itself. The magnetic field and velocity perturbation occur owing to the sudden outburst (outward acceleration) of the electron-ions due to a pulsing electrode at the FB center [4]. It is worthwhile that the direction of propagation of the Alfven wave is along the direction of the magnetic field and transverse to the direction of magnetic field perturbation [35]. It should be mentioned here that the magnetic field perturbation and this field itself are transverse to each other. Such Alfvenic phenomena naturally occur in diverse space plasmas around the Earth, laboratory plasmas, and other similar circumstances [31, 35].

### 2.6 Normal nonlinear PFS instability

There are numerous nonlinear instabilities exciting extensively in laboratory [1], space [51], and astrophysical [37] PFS systems. As already highlighted previously, besides the linear theory predictions available in the literature, there has been a great necessity for the development of quasilinear and nonlinear analyses of the PFS systems. A quasilinear model formalism to analyze the PFS instability in normal two component plasma conditions has accordingly been adopted [2]. This treatment efficiently reduces the analysis of the PFS system into a second-order nonhomogeneous ordinary differential equation (ODE) without any loss of generality. The numerical solution of the ODE is found to yield a peakon-type PFS potential structure for the first time in the literature [2].

In the recent past, the above work has also been systematically extended to implement and realize the multi-order nonlinear calculation scheme of the said PFS instability with the help of standard nonlinear perturbation methods. Similar peakonic structures are also found to result in higher-order nonlinearities. It is noticed that the peakonic structures get steeper with the increasing order of nonlinearity manifesting the unstable nature of the PFS system owing to the strong charge number density gradient [52]. The same density gradient is also found to trigger Rayleigh-Taylor instability (RTI).

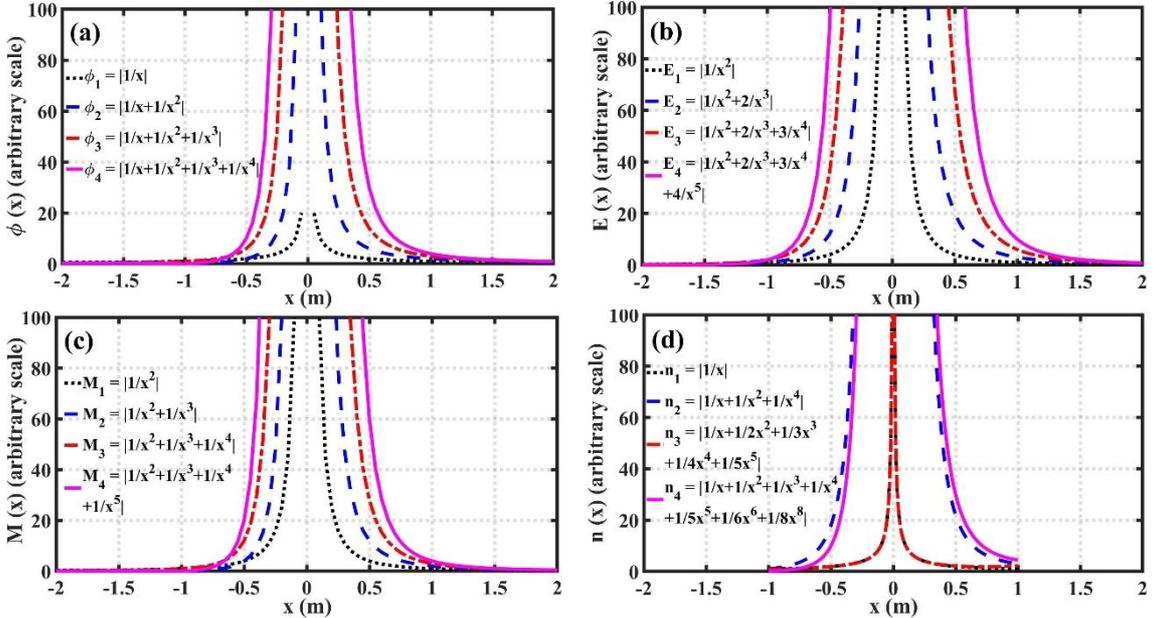

**Fig. 6** Spatial variation of relevant FB physical parameters: (a) electrostatic potential ($\phi$), (b) electric field ($E$), (c) Mach number ($M$), and (d) charge number density ($n$) associated with PFS instability (inspired by Ref. [52]).

### 3 PFS instability overview

It is already enumerated in the last section that, there are several normal plasma instabilities, excitable in a PFS system. Some of these instabilities have great significance in both pure and applied plasma science, engineering, and technology. It includes multiple technological domains including commercial material processing and manufacturing, collectively. A brief overview of the dynamics of relevant instabilities along with their various stages, requisite growth threshold conditions, practical implications, and applications together with corresponding necessary schematic diagram is presented in the following subsections.

### 3.1 Sheath plasma instability

One of the most dominant plasma instabilities excitable across the plasma sheath (also plasma FBs) is the sheath plasma instability (SPI). An externally biased electrode submerged in Maxwellian plasma in a plasma chamber is found to yield high-frequency oscillations with frequency of sheath plasma antiresonance order. The range of the oscillation frequency can extend up to that of the electron plasma frequency ($\omega_{pe} \sim 10^{12}$ Hz). This phenomenon can be explained with the help of the concept of negative resistance related to the duration of the electron drifting across the sheath with nonzero inertia. The various harmonics of instability are found to be radiating nonlinearly as EM wave pulses. These wave pulses are different from electron orbital (cyclotron) instability. It is pertinent to add further that the SPI gets excited in the sheath region and can also convectively migrate to the IFB center, thereby creating a nonzero and larger central instability amplitude. This may be due to the superposition of the SPI disturbance reaching the IFB center at same phase from the boundary sheath like a constructive (resonant) interference pattern [11]. Two specialized schematics manifesting the SPI excitation within a hollow and meshed IFB anode along with its quasilinearly perturbed potential profile are shown in Fig. 6. The standing wave like behavior of the potential profile during the SPI excitation is proved here.

The well-established capacitor (C)- and inductor (L)-like behaviors of an electron striped plasma sheath and field-free ambient plasma respectively also help in developing an idea about the dynamics of the SPI [53]. The plasma sheath-based LC circuit leads to some resonance- and anti-resonance-like conditions during the operation of current collection from the ambient plasma through the plasma sheath. The sudden temporal fluctuations in the electric current collected with varying electrode voltage insinuates the active SPI. A series and parallel resonance and anti-resonance like situations are noticed to arise around an ion depleted sheath which lead to an overall unstable behavior in the plasma sheath. This specific instability is designated as the SPI, and it is known to arise due to a negative differential radiofrequency (RF) sheath resistance of the electron rich sheath enveloping the anode volumetrically. The nonzero electron transit time through the sheath creates a phase difference of π radian between the collected current and sheath potential. This consistent phase difference is responsible for LC resonance [11].

The plasma sheath stores adequate free energy to oscillate and form a resonance with the ambient plasma (in RFB) or trapped plasma (in IFB) upon reaching certain electrostatic potential of the anode. This resonance is termed as sheath plasma resonance (SPR) and the instability is also called the instability of SPR. The frequency of the resonance is of the order of electron plasma frequency ($\omega_{pe}$). The $\omega_{pe}$ is nearly close to the inverse of electron transit time across the sheath. For an unmagnetized Argon plasma of density $\sim 10^{17}$ m$^{-3}$, thermal energy $\sim 0.5$ eV, pressure $\sim 3 \times 10^{-3}$ Torr, the SPI is found to occur at a potential biasing of nearly 60 V for an RFB like arrangement. The SPI forms spatially varying evanescent wave structures across the sheath and its amplitude varies temporally [11]. The whole PFS behaves as a single spherical shell encompassing the crests and troughs across it. These crests and troughs do not displace spatially and continue to form at their respective locations with no significant unidirectional propagation of energy through the sheath. The magnitude of their lengths only varies temporally.

The SPI dynamic within an IFB is a little different in terms of the presence of resonating plasma. Unlike an RFB arrangement, the plasma is trapped within the anode in an IFB. The plasma resonates with the sheath upon fulfilling certain parametric conditions, which vary with the host plasma. The evanescent and standing wave-like structures are formed inside the IFB anode herein (Fig. 6(b)). The growth and decay of the wave structures occur temporally [54].

The knowledge of SPI is very crucial in interpreting antenna signals received from the spacecrafts which get charged due to their movement through the ionizing gas medium. The sheath developed around the antenna may get one hazardously misinterpret the received signal. Moreover, it provides an alternative explanation of the MW emitting orbitron devices and for the negative mass instability. The laboratory and astrophysical relevance of the SPI can be well-realized from the diverse applications as above. The knowledge of the SPI can be utilized for the fundamental, technological, and industrial development leading to various emerging areas of plasma research and application [11, 54].

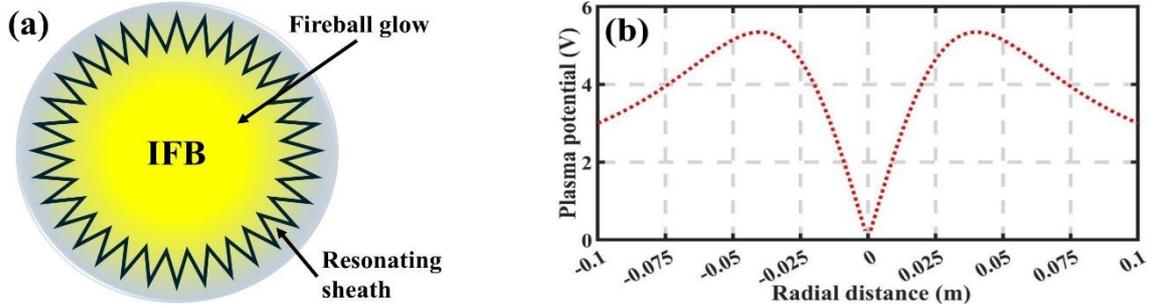

**Fig. 7** Schematic diagram showing (a) SPI excited in an IFB system and (b) its spatial potential variation. The exterior sphere (grey) is a pulsed anode. The star (black) is an SPI induced oscillating sheath. The trapped plasma IFB glow is depicted with central yellow zone in accordance with Ref. [54].

Fig. 7(a) shows a resonating IFB sheath undergoing SPI and Fig. 5(b) shows the radial variation of the consequent potential perturbation. The potential perturbation replicates that of a standing wave formed within the sheath without a propagation outside it. The nodes and antinodes formed within the sheath are radially invariant and continue to form at same radial distances under the assumed spherical symmetry. The scale of potential perturbation typically is ~10% (6 V) of the applied electrode potential (60 V) in normal plasma configurations [54].

### 3.2 Streaming instability

It is worth mentioning that one of the most naturalistic instability processes occurring in plasmas because of its collective degrees of freedom is the excitation of streaming instability, multi-stream instability, or the Buneman instability, triggered by virtue of inter-constitutive relative motions, e.g., electrons and ions drifting relative to the plasma sheath boundary. The instability threshold requires the plasma relative flow to reach the order of bulk fluid acoustic speed. An interaction of the charged particles and a wave generated due to some external influence in the same medium may lead to the growth of this instability with sufficient free energy available (particle velocity in this case). Steaming instability may develop due to several conditions. The interaction of two counter-streaming (anti-parallel) charged particle beams of comparable densities may be the simplest case as an instant example to comprehend in this instability context [36]. A simple outset of streaming instability with two oppositely facing streams of fluids is shown in Fig. 8. This simplest case of the streaming instability, termed as two-stream instability (TSI), may arise with a plane anode as shown in Fig. 13. The two required streams of fluids are produced by the anti-drifting electrons (pulled by the anode) and ions (pushed by the anode).

TSI is a special class of instabilities where the beam component of the drifting charges or ions is the free energy source for the instability onset and its subsequent growth. It may further be seen that the TSI onset threshold is given by the inter-beam (inter-constitutive) relative velocity comparable to the bulk acoustic flow of the plasma [35]. The dispersion relation derived through the linear normal mode analysis suggests a successive growth pattern of the instability amplitude in the direction of acoustic wave propagation of our concern [54].

The physical explanation of TSI can be given as the KE transfers from the electron to the ion stream, which acts as a primary agent for its instigation. As usual, the natural frequency of electron oscillation (here, in electron fluid stream) is higher than that of the ions (in ion fluid stream). But due to the Doppler shift both the electronic and ionic components may be construed to oscillate at the same frequency at certain relative inter-component velocity. Consequently, the electrons in its stream are evaluated to have a reduced KE during instability (against the previous stable states). However, the TSI still continues to grow with the KE of gross plasma fluid remaining the same. As an application of a PFS system undergoing TSI, it may be added that the TSI can generate MW radiation within an appropriate plasma system. The electron and ion bunches created due to the TSI can resonate with the resonating plasma chamber leading to the emission of MW radiation in the considered plasma system [35].

Although the quantitative features of the TSI may vary with the electrodes of different geometrical shapes (also in PFS systems), the overall physical insights behind its formation remain the same. It may be important to add here that the presence of potential relaxation instability (PRI), as being discussed in section 3.7, can change the electrode polarity, hence the direction of electric field from (or towards) it. It disrupts the ongoing streaming instability, forcing it to grow over with streams reverting their directions across the applied electrode, and so forth.

As already indicated, TSI arises both in laboratory plasmas as well as in astrocosmic environments [51]. In addition, the TSI has also been found to hold good in diverse micro-electro-mechanical devices [55], quantum viscoelastic dusty plasmas [56], protoplanetary disks [57], etc.

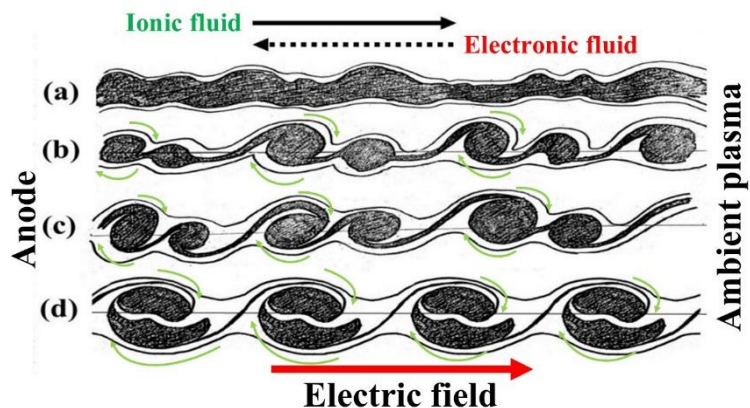

**Fig. 8** An artistic orchestration of streaming instability excited at the junction (darkened region) of two counter-moving fluids with a non-zero relative shear velocity. The two arrows (solid and dashed) appearing here denote polarity-dictated directions of counter-moving constitutive fluids (electronic and ionic). The green curved arrows show directions of spirals forming due to fluid mixing. The numbering (a), (b), (c) and (d) indicate successive temporal stages of the instability.

It is important to add that streaming instability excites several natural scenarios apart from the laboratory PFS system. Some of the environments with naturally occurring streaming instability are discussed below.
(a) Mars: It is believed that Phobos (moon of Mars) ejects dust from its body. The ejected dust forms the dust belt around the planet. The dust grains have typical physical parameters, such as, dust mass $(m_d) \sim 1.66 \times (10^{-20} - 10^{-24})$ kg, dust velocity $(v_d)$~$2.64 \times 10^3$ ms$^{-1}$, temperature $(k_B T)$~0.1 eV. These parametric values are suitable for the excitation of streaming instability [58].
(b) Jovian ring: Quasi-periodic (period~28 days) stream of dusts with $m_d$~$1.1 - 1.6 \times 10^{-16}$ kg, $v_d$~$2 - 5.6$ ms$^{-1}$ of submicron size which were detected by mission 'Vlyses' to Jupiter. The instability onset conditions are fulfilled by these dust particles to excite the streaming instability in the Jovian ring [58].
(c) Comets: The typical values of the dust parameters are reported to be $m_d$~$10^{-19}$ kg, $v_d$~$4 \times 10^5$ ms$^{-1}$, $n_d$~$10^3$ m$^{-3}$, and $k_B T = 0.1$ eV. Thus, the cometary environment is plausible for streaming instability excitation [58].
(d) Interstellar media (ISM): The shock waves observed in the ISM beside the star forming regions yield $v_d$~$10^4$ ms$^{-1}$, $T$~$10 - 20$ K, which are feasible parametric values for the streaming instability formation [58].

Thus, it may be concluded that studying streaming instability under various laboratory plasma conditions (e.g., in PFS system) may provide a simulative platform for studying the same in actual natural scenarios. This proves the relevance of instability studies in specially arranged PFS systems.

### 3.3 Secondary ionization instability

The plasma medium constituent electrons experience an incessant drag towards (or away from) the submerged electrode (in a PFS system) due to its pervading electric field. The field-drifted electrons while moving across the plasma, often collide inelastically with the constituent neutrals, liberating valance electrons therefrom. The primary electrons liberated as a direct consequence of the primary ionization acquire more kinetic (free) energy from the background electric field sourced at the installed electrode. The accelerated primary electrons may again knock out more (other) electrons from the constitutive plasma ions as well as fresh neutrals resulting from recombination processes, thereby leading to a phenomenon termed as the secondary ionization instability (SII).

It is known that electron beams with periodic electron bunches can instigate plasma oscillations in the medium. It may be added here that the bunches in the electron beam are not required throughout to instigate plasma oscillations. But, once the sheath oscillation is set up, any constant flow of electrons can also form bunches within themselves as a self-sustained feedback mechanism [35]. As a result, a constant voltage PFS electrode can also lead to the excitation of plasma waves. Thus, the electron bunching and ion acoustic wave (IAW) developing a SII is an experimentally imminent observation [4]. An artistic orchestration of possible IAW during the formation of a PFS in a plasma chamber is shown in Fig. 9. The darkened (denser) and less darkened (less dense) regions show crests and troughs of the IAW formed. These two different regions have electrons and ions with alternating magnitudes of KEs. The direction of the IAW is pointed outwards from the anode towards the ambient plasma and its wavelength also increases (energy decreases) while moving outwards. With the increase in secondary ionization the wavelength also varies temporally as the sheath shielding the anode electric field thickens.

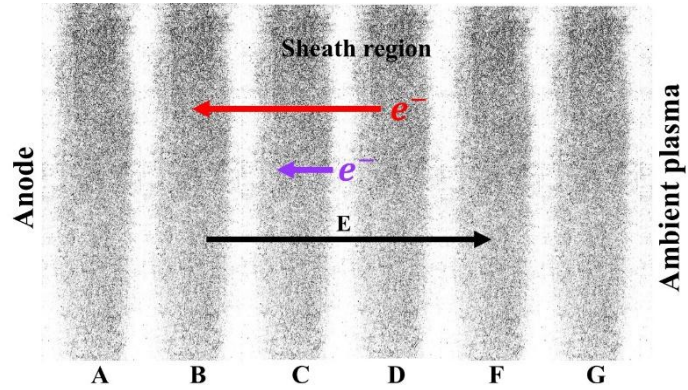

**Fig. 9** Schematic diagram showing electron density growth due to IAWs in circumvent PFS region. Compressed regions over alphabets, viz. A, B, C, D, F, G have a higher electron population density than those in between [35]. Red and violet arrows denote faster (from crest) and slower (from trough) electrons, respectively. The largest arrow (black in color) points towards the direction of sheath electric field.

It is important to note that some longitudinal low-frequency oscillations (frequency ~ $5 \times 10^2$-$2 \times 10^3$ Hz) can also be detected across the DLs around the PFS. The adjacent layer of sheath can grow the IAW, which propagates via contraction and rarefaction in the plasma medium, triggered by density perturbations. Due to the growth of contraction and rarefaction associated with the longitudinal IAWs, the electrons in the contraction region acquire unidirectional KE towards the anode. The electrons having been adequately energized at the high-energy IAW threshold magnitude may knock out other electrons from the neutrals (with inelastic collisions) present in the medium leading to the excitation of a SII. The electrons knocked out therefrom may further feedback the ionization processes repeating the same energization and collisions. In a double plasma device (DPD), there are enough electrons across the DLs with energy higher than the ionization potential of the host gas. It is seen that both the IAI and the SII are inextricably linked because of the intrinsic fact that one can trigger the other [35].

It may be interesting to note here that the same bifluidic plasma model, as already used in the TSI study, also holds good for triggering the ionization instability. The instability grows only upon having a greater cross-section for collision with ions rapid enough to overpower the damping due to the electron-ion recoupling, viscosity, and the soft collisions themselves. In a nutshell, the required threshold conditions for the onset of a consistent ionization instability can be enlisted as follows:

(a) For a requisite collisional cross-section, the density of the gas, and hence, its pressure is supposed to be at least $2 \times 10^{-3}$ Torr.
(b) The frequency of the instability is measured to be a few kilohertz and wavelength equivalent to the length of the system. The dimension of the system shrinks the energy of oscillation into wavelengths of the same order as that of the system length.
(c) The strength of the instability is gas specific. For example, Xenon, Krypton, and Argon gases are observed to be suitable for the excitation of the instability, but the instability is never observed in the Neon or Helium gases. This must be due to the different ionization energies of the respective gases, which yield different electron-ion pair production, even at the same collisional cross-section [60, 61].

For the astrophysical significance of SII in detail, one may refer to the significant work by Bagińska and his team [62]. A thorough study of this instability through the laboratory PFS system with pertinence may yield efficacious results in diversified astrophysical terrains also [63].

## 3.4 Ion-acoustic instability

It is well known that an electric field pervading a bifluidic uniform plasma medium may yield low-frequency (in Hz) IAWs across it. The IAWs grow due to relative streams of charges driven by the electric field produced by any externally biased grid or electrode (in this case) submerged in the system. The field-drifted charges due to their inevitable collisions with neutrals (less in number) and consequent ionizations form lumps of charges. These lumps take form of a longitudinally propagating inhomogeneous density-gradient driven plasma instability, termed as ion-acoustic instability (IAI). The IAI in appropriate conditions propagates with consecutive high-density (compression) and low-density (rarefaction) regions in a PFS system to the ambient plasma from the anode. The densities of two consecutive rarefaction and compression regions do not remain the same but vary uniformly and gradually while approaching the electrode [63, 64]. In the linear perturbation scheme the waves with temporally increased amplitude, are termed as instability. Thus, one of the fundamental differences between IAW and IAI is that the latter has a gradually increasing amplitude, ideally reaching infinity at infinite time. A simplified artistic representation of the IAI is presented through a schematic in Fig. 10.

The relative shear flow of charges with varying number density and electron temperature anisotropy ($T_e^\perp \neq T_e^\parallel$) develops oscillatory features (longitudinal) leading to an IAI. It is added that, even with a meagre frequency magnitude, the oscillation of instability possesses an amplitude higher enough to skip wave damping. This damping introduces stability to the system. The IAI has a wide range of realistic applicability, such as in solar wind, Earth's bow shock, magnetotail, field lines of aurora in planetary magnetospheres [65], *F*-region of the ionosphere [34], and so on.

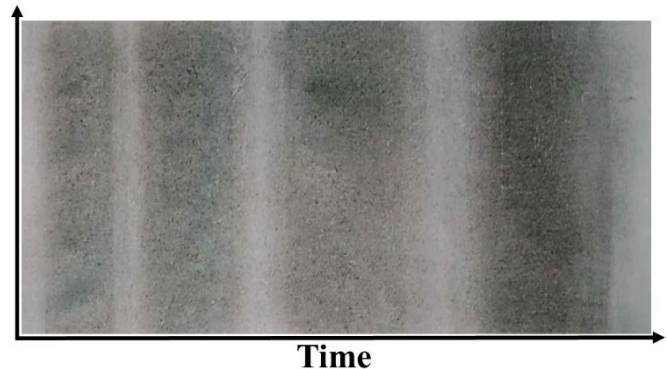

**Fig. 10** Schematic diagram showing a visualization of IAI at certain fixed spatial distance from the electrode. The increasing amplitudes (widths) of dense (darker) and less-dense (less-dark) areas with respect to time insinuate the IAI excitation in a PFS system.

It is worth mentioning that the IAI triggers the formation of a normal DL, which is paramount for the formation of the FB glow in an adequately large plasma-electrode system, as per the local Bohm criterion [65]. The anomalous resistivity generated as an outcome of the IAI builds up an electrostatic potential equivalent to the ionization potential of the host plasma gas. These bipolar radially or axially symmetric, highly nonlinear regions of electrostatic potential, are termed as DLs. It is important to add that the DL is very necessary in a PFS system for accelerating the electrons with energies high enough to undergo inelastic collisions with other plasma constituents. To form the DL and hence PFS glow, the field induced drift ($v_{de}$) speed of the electrons must be greater than their thermal speed ($v_{te}$). It may be noteworthy to add that the DL and TSI can excite each other in a laboratory PFS system.

On the contrary, in auroras, along magnetic field lines, the weak magnetic field may result in a violation of the velocity inequality condition ($v_{de} > v_{te}$) to excite the IAWs at $v_{de}$ ($< v_{te}$) also. However, the TSI cannot be excited there while having a subcritical electron drift speed ($v_{de}$), lower than the electron thermal speed ($v_{te}$) [66]. In other words, it is, hence, inferred that the TSI excitation is not possible due to subcritical conditions for its onset in the system.

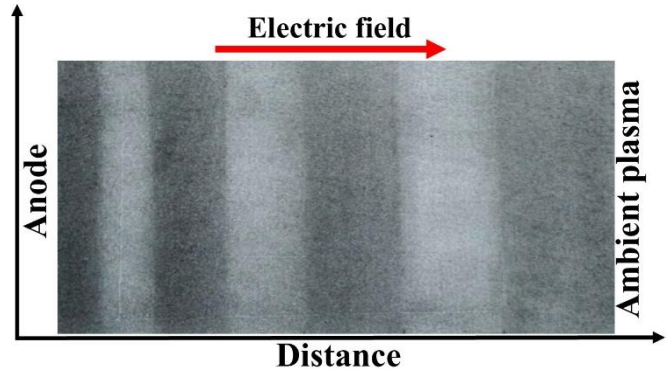

**Fig. 11** Schematic diagram showing a visualization of IAW at certain asymptotic times. Alternate denser (darker) regions gradually fading to less-denser (less-darker) regions denote decaying number density and its spatially increasing wavelength with respect to distance. Denser regions are separated by rarer (less-darker) regions in between, which also gradually widen over distance.

It should be emphasized here that there are certain differences as well as similarities in the IAW and IAI patterns. In the first case, the amplitude remains constant over time, whereas, in the latter, the amplitude grows temporally as long as the instability behaves linearly before the saturation occurs. In contrast, both the IAW and IAI show increasing wavelengths over spatial distance. The denser (darker) regions gradually lose their higher densities and the density contrast between the denser and rarer regions diminish over distance in both the IAW and IAI structures. The figure below shows the spatial variation of denser and rarer regions in the case of an IAW.

### 3.5 Rayleigh-Taylor instability

This is one of the most significant density-dependent instabilities in fluids with an operational effective gravity. It finds its relevance in a wide-range spectrum ranging from a laboratory plasma chamber (across the plasma sheath) [67] through the volcanic eruption on the Earth's surface to the extra-terrestrial nebulae (e.g., Crab Nebula) [68]. This instability can be fundamentally categorized as an inevitable expression of potential energy between two fluids of variable density through their intermixing. The two fluids of different densities ($n_1$ and $n_2$) with nonzero density gradient across their junction mix to form a single fluid of common density ($n$). The density gradient is zero upon the mixing of the fluids. The required free energy for the onset of this instability is served by the potential energy (acquired by virtue of position) of a heavy fluid (denser) residing upon another light fluid (rarer) with an effective gravitational potential acting towards the lighter fluid. With the interaction of the two fluids, the instability begins with a minor perturbation in the adjacent layer of the two fluids [38]. This stage of instability, which is at a pseudo metastable equilibrium condition, can be expressed linearly. The pseudo equilibrium condition is due to the two fluids with individual uniform density resting upon each other with different comparative intercomponent densities, $n_1$ and $n_2$, as seen in Fig. 12. The various stages of RTI are explained in the next paragraph after Fig. 12.

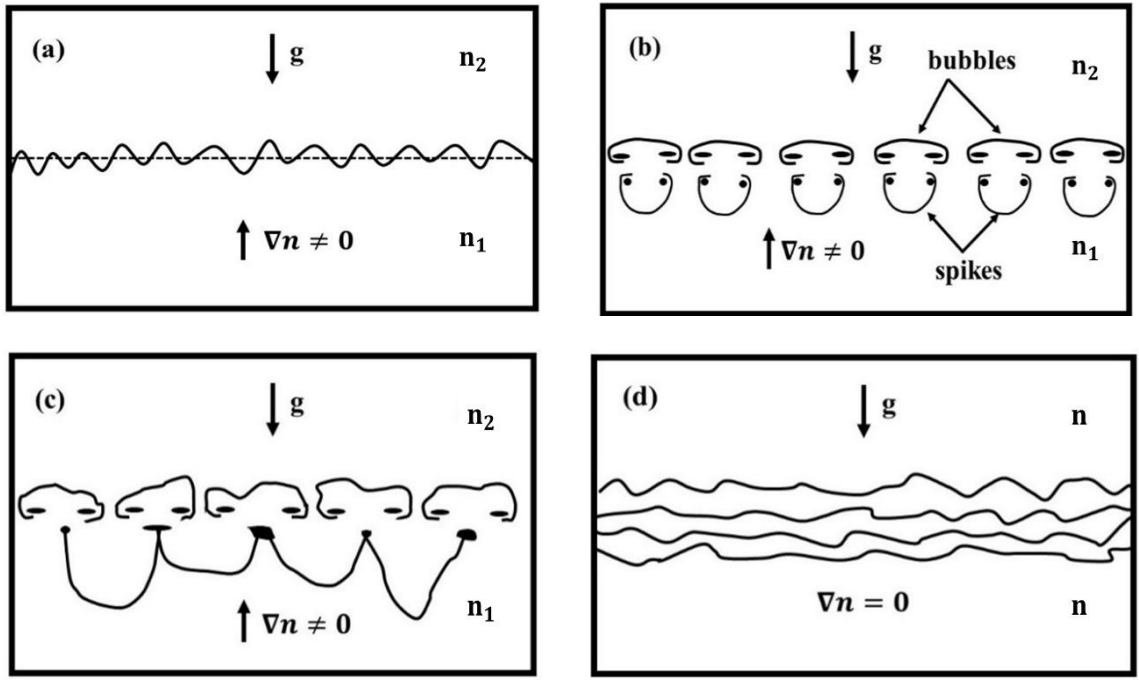

**Fig. 12** Schematic diagram showing four different stages of RTI in a PFS system. Here, ↓ $g$ and ↑ $\nabla n \neq 0$ denote direction of effective gravity and its non-zero density gradient, respectively. Effective gravity arises from the strong electric field of FB anode. Fig. 12(d) manifests final homogeneous plasma equilibrium stage with common density $n$ and density gradient, $\nabla n = 0$ [55, 38].

The growth rate of the RTI at the first stage is evidently linear in time (Fig. 12(a)) due to small amplitude perturbation. The second stage of instability (nonlinear in nature) witnesses the formation of 'spikes' and 'bubbles' due to the inter-fluidic migration of heavy and lighter fluids (Fig. 12(b)). It is noteworthy that there is no apparent mixing of the 'bubbles' and 'spikes' so far. The third stage of the instability (nonlinear) begins with the intermixing of them (Fig. 12(c)). The previously formed shapes of the 'bubbles' and 'spikes' now acquire some unrecognizable shapes (slow in change). The inter-fluidic patterns, formed previously, now get enveloped by the continuous inter-fluidic migrations. The fourth stage is the inter-fluidic turbulent mixing of them. With the fluid mixing accomplished, the final equilibrium stage of the two fluids arrives (Fig. 12(d)) little free energy remaining for any inter-fluidic movement. Consequently, the fluids arrive at a stable equilibrium with no possible particle dynamics without an external influence [38].

In a PFS system of spherical (Figs. 3-4) or planar geometry (Fig. 13), this RTI may be instigated as the necessary metastable quasi-equilibrium density condition is satisfied between the sheath around the anode and the ambient plasma. Moreover, the effective gravity in this arrangement is produced by the sufficiently strong anode electric field. The strong anode electric field in a PFS behaves as the effective gravity for the charged laboratory plasmic constituents just like actual gravity behaves with the massive astrophysical charged as well as neutral fluids. The anode is enveloped by the plasma components forming the sheath which has a strong internal electric field across it, which acts as the effective gravity. This RTI can occur simultaneously with the potential relaxation instability (PRI) (being discussed in section 3.7) due to weakening anode potential, and the consequent density gradient can instigate the RTI. The outcome of RTI, along with that of the PRI, causes the destruction of the sheath [40], which again reforms due to the consistent electrostatic potential across the electrode from the external biasing in the system.

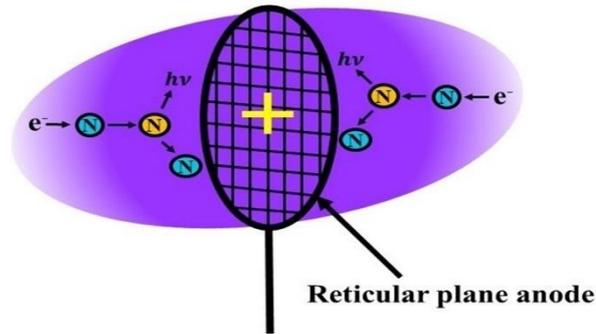

**Fig. 13** Schematic diagram showing plane electrode (anode) FB formation with all usual symbolic indications described clearly in main text (inspired by Ref. [10]).

The phenomena of excitation, ionization, and energy release through the emission of visible light along with dominant instabilities are the same here as that of the spherical (solid or reticular) anode as discussed above. The electrode frame is made up of some solid material and the reticular plane is made up of adequately thick conducting material. The electrons collide with neutral atoms in their ground states (shown as bluish regions) and excite them to higher electrostatic potential (shown as yellow regions) [1, 8]. The extra energy ($h\nu$, $h$: Planck constant, $\nu$: frequency) is successively released from the excited neutrals and it forms the PFS glow around the biased plane anode [4].

Apart from the laboratory plasma chamber, the RTI is known to excite in several other astrophysical environments. In the Sun, the plumes observed are an outcome of the RTI excitation caused by the heavier plasma filled solar prominences lying atop the lighter coronal region. The prominent region is two orders of magnitude denser and cooler than the surrounding coronal region. The RTI excitation creates characteristic mushroom like plumes [69]. Besides, the nebular structural formation is also known to occur through the RTI mechanism. It is proven through the observation that RTI plays a role not only in shaping the outskirts of the nebulae but also acts to shape the mechanism which allows ISM to flow inside the nebular inner parts by making halo fragments [70]. The common dynamics of RTI in both PFS and other astrophysical events prove the relevance of RTI study through the simpler PFS system.

### 3.6 Kelvin-Helmholtz instability

The Kelvin-Helmholtz instability (KHI) or shear instability is an omnipresent dynamic instability phenomenon observed in diversified fluidic environments in different physical conditions. It pervades terrains ranging from 500 meter deep beneath the water surface (metalimnion), through laboratory plasma chambers on the Earth (e.g., DPD), clouds in Earth's atmosphere (Fig. 16), up to interstellar dust molecular clouds (e.g., Orion) in astrophysical domains [37].

The KHI is an outcome of nonzero relative flow motion (shear velocity) between two layers of same or different stratified fluids [37]. It may be added here that the necessary shear among the fluid layers can originate due to the intrinsic viscous property of a fluid and not necessarily due to one faster fluid imposed upon another stationary or slower fluid. The interaction of two (or more) fluid layers, which appear to be linear at the beginning, starts with sinusoidal ripples formed at their junction interface (Fig. 15). But such spiral patterns take purely nonlinear structural forms as time passes on. The small amplitude apparently regular sinusoidal ripples while moving ahead, gradually develops into a string of irregular shapes. With the subsequent shifting from the linear to nonlinear regime, the fluidic ripples successively form enlarging spirals (Fig. 15). The spirals further bifurcate as a duo of massive bodies revolving around a common center of mass with a bunch of disturbances emanating outwards [71].

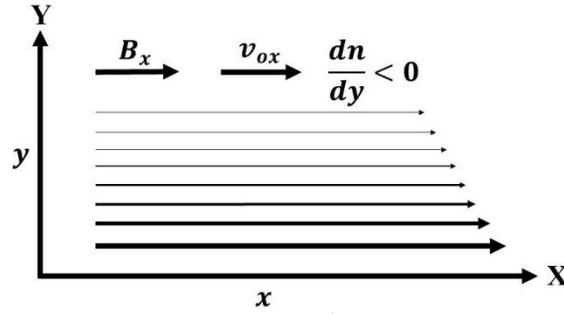

**Fig. 14** Schematic diagram showing directions of various active parameters during KHI onset. The gradually thickening-up x-pointing arrows show the increasing density gradient trend towards negative $y$-axis [72]. Lengthening arrows downwards insinuate at increasing velocity of different fluid layers. $B_x$, $v_{ox}$, and $dn/dy < 0$ denote magnetic field (along x), unperturbed velocity (along x), and density gradient (downwards along y), respectively.

It is noteworthy that both the density and velocity gradients between (or among) the layers (Fig. 14) of the viscous fluids inherently bolster the expression of free energy due to the inter-layer velocity or density inequality. The free energy remains stored among the fluids through their density inhomogeneities without assimilation upon preliminary interaction. The free energy is expressed at the assimilation of the fluids yielding a collective common homogenous density. This instability is more dominant in plasma chambers, especially with externally biased electrodes, as they can create a shear among plasma fluids of variable densities [73]. For example, a magnetized PFS system with inhomogeneous plasma density can yield a shear among the various layers of plasmic fluids. Besides, in the cubical IFB arrangements with the anode faces biased to different voltage [74] can create a shear among the various layers of the plasmic fluids creating a suitable condition for KHI excitation. Thus, the KHI excitation may start from the anode extending up to the ambient plasma through the sheath in between.

The temporal fluctuations of the electric field maintaining the inter-layer density variation may trigger the inter-mixing of these layers. A situation of this kind arises in the PRI discussed in the next section (section 3.7). Besides, there are unidirectional flows (streams) of electrons and ions between the submerged electrode and ambient plasma yielding an uprise of the KHI in the PFS system.

In the PFS system under consideration, it is emphasized herein that any electrode of plane or spherical geometry can excite the KHI [72]. The outcome of a systematic study of the KHI through a PFS system can be extrapolated and mapped into atmospheric physics also. The multiple layers of clouds with variable densities are proved to undergo the same spiraling motions as mentioned previously (Fig. 15), highlighting its relevance in the study of turbulence for weather forecast, and so forth. The sheath and ambient plasmas resemble that of multiple layers of clouds and may act identically witnessing the same RTI and KHI [75]. For further KHI studies in laboratory produced dusty plasmas, one may refer to the work of Kumar and his team [76]. A few conditions for the KHI to get excited in a PFS system are explained as follows:

(a) In the case of conducting incompressible fluid moving in the direction of the applied magnetic field, the KHI occurs only if the relative speed between two adjacent fluid layers is greater than the root mean square Alfven speed of the layers (Fig. 14). For compressible fluid, the velocities of the individual fluid layers must at least be equal to the propagation speed of the perturbations in the fluid media.
(b) For neutral (non-plasmic) fluids in a gravitational force field (without magnetic field), the KHI onset condition derived by Chandrasekhar in a usual physical symbolism is given as: $(\partial v_{oz}/\partial x)^2 > 4g|1/n(dn/dx)|$. Here, $v_{oz}$ is the fluid velocity along the z-direction (orthogonal to x-axis), $g$ is the acceleration due to gravity at the Earth's surface, and $n$ is the fluid particle density [72].

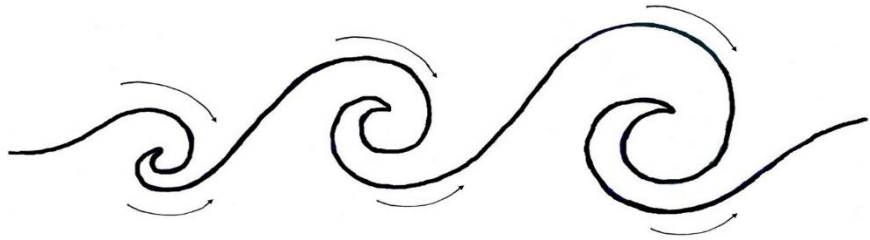

**Fig. 15** Schematic diagram showing KHI evolving between two fluid media of different densities with the rest explained in main text in detail. Curved arrows show direction of one fluid flowing (mixing) through (with) another fluid.

The continuous spirals in Fig. 15 denote the denser medium migrating through and then mixing with the rarer ambient medium. As evident, the KHI is linear (apparently sinusoidal) in nature at the beginning. However, it gradually passes through a quasilinear phase until the subsequent nonlinear phases (spirals) have arrived. The first (sinusoidal) phase can be formulated mathematically using linear formalism. When the rate of change of a physical parameter is directly proportional to its instantaneous magnitude [54]. In successive phases the two media form more spirals with widening area. Eventually, the largest single spiral divides itself into two; but, of dimension, apparently smaller than the former one. This phase must be dealt with lower-order nonlinear formalism. The dense spiraling fluid media are gradually noticed to fade while migrating through the ambient fluid media ahead. The spirals behave as if spinning around some massive objects at their center emanating substances outwards. The inter-component mixing ahead gradually acquires some irregular shapes and fades into the ambient media. This inter-fluid coupled stage is purely nonlinear in nature, possessing variable and complex rate of change to be dealt with a nonlinear mathematical fluid dynamics treatment [77].

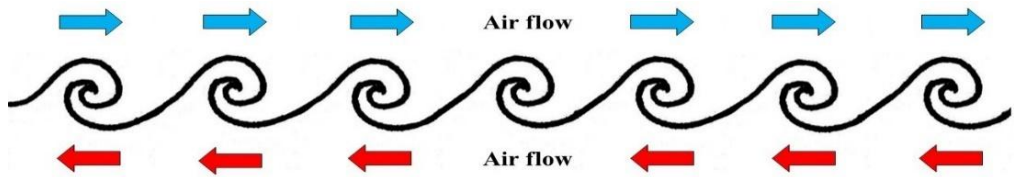

**Fig. 16** Schematic diagram showing intermixing of water vapors (in middle) and air media (enveloping central water vapors) in clouds saturating nonlinear structures via KHI excitation processes with the rest detailed in main text.

As discussed above for charged fluids inside a plasma chamber, the similar velocity shear and density difference across the stratified fluid layers can also naturally occur in some non-plasmic scenarios resulting in the excitation of KHI. Two of the commonly observed KHI excitations are discussed herein with refence to the KHI dynamics illustrated before. The various layers of the clouds can have velocity shear (due to different wind speeds at different altitudes in the atmosphere) and density difference (due to Earth's gravity). The faster (lighter) layer of cloud scoops the top of the slower (denser) layer below forming the asymmetric wave spirals pointing towards the direction of the faster layer [78].

Akin to the clouds, the stratified layers of sea water also have velocity shear with density differences. The velocity-shear originates due to the touching wind flowing above the sea surface and intrinsic water viscosity. The water layers beneath behave as stationary against the faster wind leading to the formation of the KHI on the sea surface. It must be highlighted that wind itself also behaves as the topmost layer in the KHI excitation on the sea surface. Same as the clouds, the vorticity, roll-up, and wavelike shapes also form in sea water [79]. The source of free energy in both scenarios for the KHI excitation is provided by the shear among the different fluid layers. The required shear in the velocity in PFS for KHI excitation is provided by the inhomogeneous cubic anode biasing [74]. Thus, the PFS induced KHI study is useful in simulative exploration of Earth and spaces sciences.

### 3.7 Potential relaxation instability

The potential relaxation instability (PRI) has a purely nonlinear character, which excites due to a plasma submerged electrode with electrostatic potential higher than the plasma potential ($> 50$ V). The PRI can be defined as the unstable plasma situation in an anode-peripheral region across a magnetized PFS system caused by the combined coordination of the constitutive number densities and charge fluctuations. The free energy required for the onset of the PRI here originates from the externally applied electrode biasing in the PFS system. This instability can be identified in the form of an array of large periodic fluctuations ($> 50\%$) in the plasma potential, electrode current, charge number density across the central electrode. This instability also acts as a radially (for spherical electrode) or axially (for plane electrode) moving DL. This is due to the spatially steep nonlinear parametric variation of instability. The existing distance of the PRI depends on the ratio of the parallel ($D_\parallel$) and perpendicular ($D_\perp$) electron diffusion coefficients (relative to the applied magnetic field) along with the electrode width [40, 80]. This PRI dynamics across a plasma submerged anode can be illustrated as follows.

The PRI triggers as soon as the electrode potential ($\phi_E$) exceeds the plasma potential ($\phi_p$). Increasing the $\phi_E$ further increases the instability frequency and amplitude until acquiring a maximum at $\phi_E - \phi_p > 1$ V. Increasing the $\phi_E$ further ahead reduces the amplitude but the frequency continues to grow. The frequency and the amplitude of the PRI are measured by the spectrum analyser. The PRI is also manifested in the form of a moving DL across the plasma chamber perpendicular to the electrode surface, towards or away from it. In other words, DL shows a back-and-forth motion during the PRI excitation. The decay in the electrode current noticed during the PRI is due to the potential depression on the lower potential side of the DL. The accompanying coherent collector current oscillations proves the PRI excitation [80]. The various stages of the PRI may be further explained illustrating the individual stages as follows.

Once the electrode biasing is increased beyond the plasma potential (Fig. 17(a)), the electrons start to gather around it and the ions get repelled back from it (Fig. 17(b)). In no time, the accumulated peripheral electrons are absorbed by the anode until an electron current saturation stage is reached. The anode having absorbed the peripheral electrons temporarily loses its positive polarity and acquires a neutral stage (Fig. 17(c)). In the absence of any substantial electrostatic force field from the neutral electrode, both the electrons and ions from ambient plasmas homogenously gather around it again (Fig. 17(d)). Due to consistent external biasing, the electrode soon acquires its positive polarity to start over the same procedure again. These dynamics of charge absorption and repulsion, leading to a massive charge density fluctuation ($> 50\%$) are termed as the PRI. The frequency of PRI is reciprocal of the temporal duration between two consecutive current or density spikes noticed in the PFS system [40].

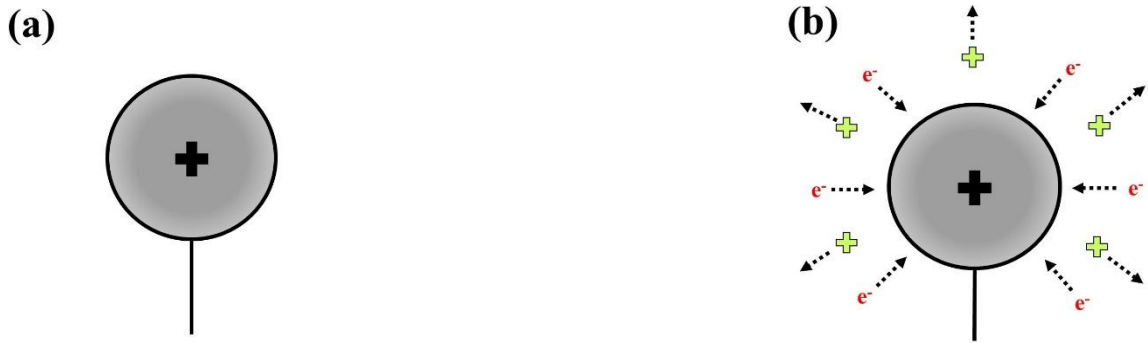

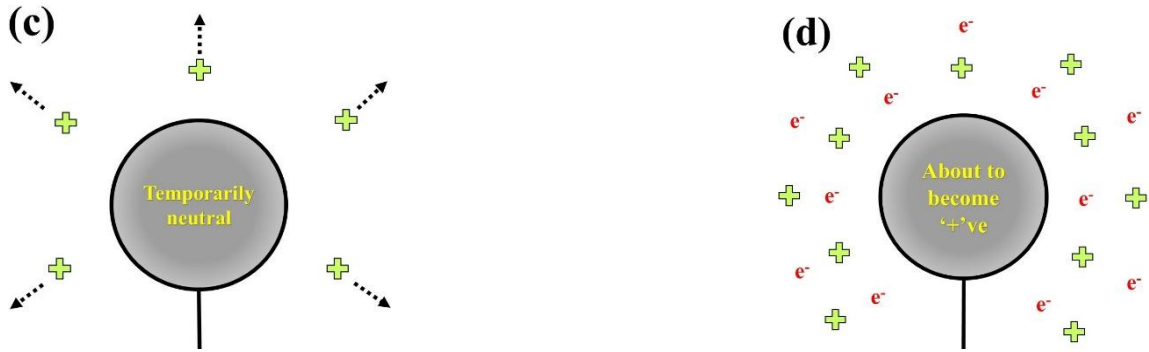

**Fig. 17** Distinct phases of PRI may be pictorially illustrated with the help of four distinct steps: (a), (b), (c), and (d). These distinct steps designate four stages of PRI. PRI dynamics start over after the system accomplishes step (d).

As shown in Fig. 18, a schematic sketch is drawn to depict an anticipated temporal variation of (a) electric current collected through the anode and (b) electric charge density fluctuations in the PFS system. Here, Fig. 18 (a) manifests two cycles of electron current collected through the anode during the potential fluctuations in the PRI. The sections OA and BC denote the second stage of instability (Fig. 18(a)); when electron gathering around the anode and its absorption reaches the peak value. The sections, AB and CD, denote the successive falls of the electron current absorption because of reduced potential value of the anode. The same cycle repeats again as shown by points B, C, D, and so forth. Fig. 18(b) shows the anticipated charge (electron and ion) density fluctuations around the anode. There is an equilibrium charge density ($n_o$) around the anode initially. Due to the growing positive polarity of the anode, the electron (ion) population density, $n_{e(i)}$, rises (falls). With a fall in the electron absorption and anode potential, the ions drift back to an equilibrium population density distribution.

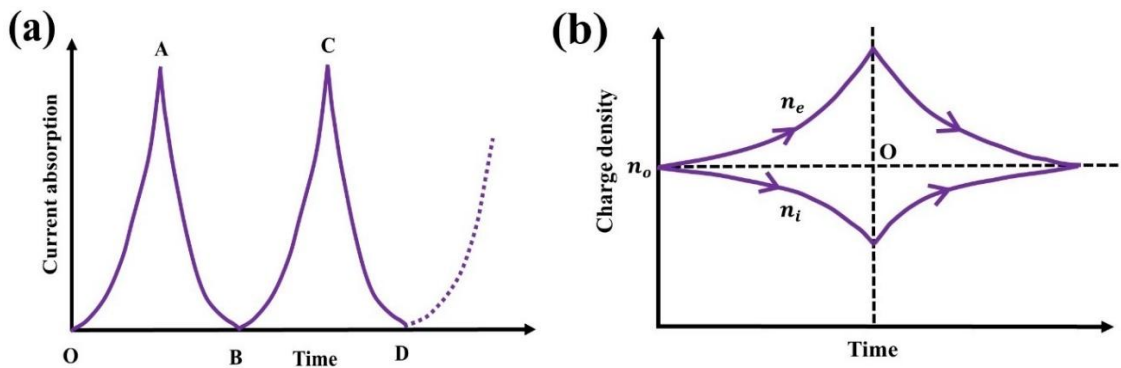

**Fig. 18** Schematic diagram depicting an anticipated temporal variation of (a) electric current collected through anode and (b) electric charge density variations in PFS system.

As discussed above, there are numerous pertinent plasma instabilities which have been experimentally noticed to get excited in the PFS system and their investigation is still being carried out. Comprehensive studies of these instabilities embolden the basic understanding not only about the plasma media, the oceanic terrains, weather sciences (atmospheric physics), as well as in diverse astrophysical scenarios. Having thrown some light to the pertinent PFS instabilities, we want to discuss some instability damping mechanisms. These mechanisms enable the PFS system to stabilize against excited instabilities, thereby leading the plasma system towards a stable equilibrium.

## 4 Damping of PFS instabilities

The knowledge of the PFSs and associated instabilities are very important in sheath physics from the point of view of both fundamental as well as applied research. Besides, a precise idea of the damping prevalent in the plasma media against the pervading waves and instabilities is also necessary. While analyzing the damping, it is found that in the plane wave instabilities in one dimension, the electric field does not damp exponentially as predicted by Landau [81]. Rather, it is found to undergo damping as a reciprocal power of time. The order of time parameter is dependent upon the initial conditions. It is also needed to add that only waves or instabilities having a wavelength greater than certain threshold length undergo damping. Those instabilities having scale lengths below the threshold length continue to develop therein as long as their free energy source continues in the system [82]. It is pertinent to add here that the threshold wavelength depends on both the system and the active instability. With the weakening nonlinearity of the instability, its wavelength may increase, which in turn may attenuate the disturbance amplitude. But the normal instability mode with the plasma particle speed greater than any other constitutive particle can survive at the cost of its free energy [83].

It is in parallel worthwhile that, in a fluid plasma description, inter-component collisions even with weaker frequency can considerably damp an instantaneous active plasma instability. This collisional damping may enable a non-Maxwellian plasma system undergoing instability to stabilize into a stable one beyond the intercomponent collisional rate itself [83]. This collisional damping may also result in the active excitation of electromagnetic waves having modified characteristic strengths [11]. The schematic in Fig. 19 shows plasma constituent gaining (losing) energy from (to) the perturbation wave under different circumstances, respectively. The first case leads to instability and the second leads to stability or damping.

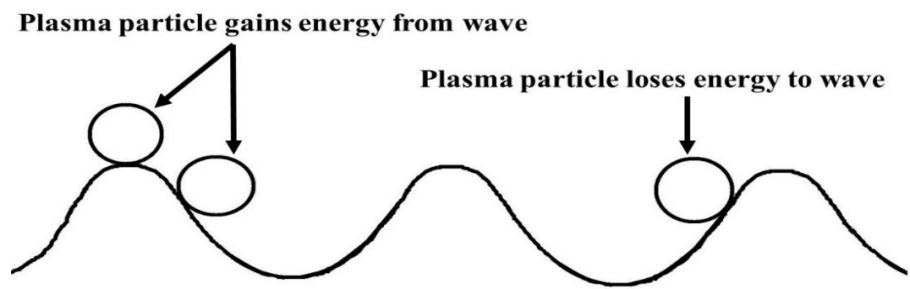

**Fig. 19** Schematic diagram showing mechanical analogy of the Landau damping mechanism. Particles slower (faster) than instability disturbance gain (lose) KE from (to) wave instability. Damping or instability in fluid medium depends upon particle speed (inspired by Ref. [35]).

A precise idea of the plasma damping mechanism learned through the PFS system finds its relevance in different branches of science and technology. One of the atypical fields of research having such wave damping relevance is the actively growing area of particle accelerator physics. The decoherence and filamentation of the accelerator beam due to the nonlinear fields is often witnessed to create confusion among the concerned researchers as the Landau damping phenomenon. This damping mechanism is necessary for smooth stabilization of particle beams in hadronic physics [84]. It finds potential applications in the kinetic treatment of galaxy formation, where stars can be considered as constitutive atoms of a plasma system, interacting via the nonlocal gravitational forces [35]. These applications advertently prove the relevance of various damping phenomena active in a PFS system, which further assists in the growth of interdisciplinary research.

## 5 Interdisciplinary technological scope

There exists a wide range of technological significance of different PFS studies and experimentations in real-world interdisciplinary circumstances. Especially, the model analyses of the pertinent excitable plasma instabilities in the PFS system find its applications in diversified interdisciplinary fields, such as nebula formation (RTI) [68], weather sciences (KHI) [77], oceanic sciences (KHI) [79], fluid dynamics (RTI, KHI), antenna signaling (SPI) [11], and so on. Besides, the sheath potential drop in regular and IFBs can be used in collecting isotopes for medical as well as atomic energy harnessing, one-step nanodot fabrication (Fig. 20(a)) [85], and pressure jet development in aviation technology (Fig. 20(b)) [80], etc.

To illustrate the above, a set of reticular spherical electrodes with customizable electrostatic potential, used in the IFBs, can be utilized to collect isotopes from a heterogenous gaseous mixture with suitable density of range $10^{15}$ to $10^{17}$ m$^{-3}$ and pressure of $5 \times 10^{-3}$ mTorr. The reticular IFB electrodes with a fluctuating electric field and suitable ion-cyclotron frequency can collect the ions and their isotopes at resonance. The collected isotopes, as per their radioactive properties, can be utilized in energy harnessing, cancer treatment, procuring defense equipment, and so forth [86].

As in Fig. 20(a), we show a plane and circular anode with applied magnetic field (to reduce contact area of the anode with ambient plasma) to align the anode flow. The ions get substituted on the GaSb substrate forming nanodots for further usage in the field of digital electronics and communication technologies. This one-step fabrication process is commercially more feasible compared to the two-step regular fabrication [85]. A stream of ions with possible pressure jet formation is displayed (Fig. 20(b)). The suspended cylindrical anode recoils towards the red arrows due to the ion thrust (Fig. 20(b)). The PFS jet thrust has potential future scope in the interdisciplinary fields of astrolabcosmic explorations, plasma-assisted technology, and aviation engineering [87].

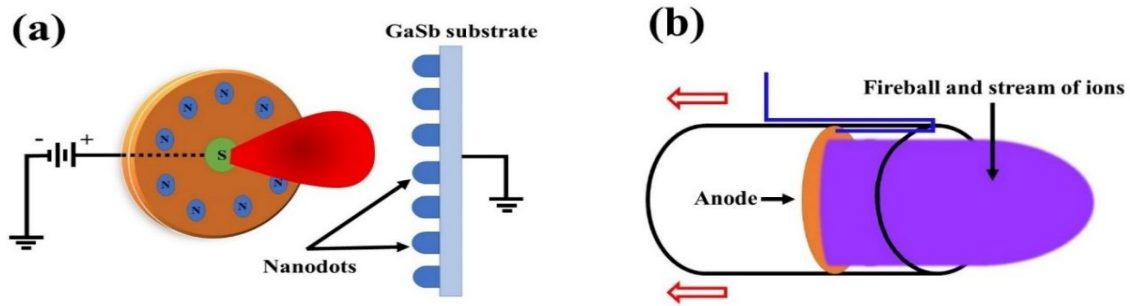

**Fig. 20** Schematics diagram showing (a) creation of nanodots (commercial one-step) and (b) stream of ions (possible jet thrust) using plane anode FB (inspired by Ref. [87]).

In addition to the various FB-centric instabilities above, we find it relevant to elaborately discuss a few of the applications of special SPI for the sake of plasma physics community. As briefly highlighted before, the SPI finds its relevance in antenna signal processing. The antennas in the spacecrafts get charged in the astroplasmic environment yielding an enveloping sheath around it like a sheath in laboratory RFBs. Consequently, the SPI can excite across the antenna at potential fluctuations. The SPI development hinders signal communication with the ground station [28]. Dealing with this situation demands knowledge of SPI, predominant in laboratory FBs. This proves the relevance of both FB research and associated instability in diverse astroplasmic scenarios. In addition, the IFB model can be used to study the magnetar (magnetic neutron star) energy emission [88] and ultra-relativistic gamma-ray burst (GRB) [89] also.

The magnetars hold very strong magnetic fields within it ranging from $10^{14}$ to $10^{15}$ G. They produce energy through internal field restructuring. The various pulses of energies produced through restructuring are either prevented by the magnetospheric sheath formed around the star or are released, depending on their magnitudes [88]. This energy release resembles an IFB sheath preventing or allowing energy pulses (depending on pulse magnitudes) during the SPI excitation.

The ultra-relativistic GRBs have another astrophysical resemblance in analogy with diverse laboratory IFBs. For example, Neutron stars and Blackholes are reported to release shockwaves at ultra-relativistic speeds. The impulsive shockwaves, while gradually moving outwards, get slower and start inter-shock interactions, like inter-pulse interactions in IFBs during SPI excitation [89]. Therefore, the IFB model during the SPI excitation processes is also useful in studying the GRBs proving the relevance of both IFB and SPI studies.

In addition to the above, FBs can also be extensively studied in quantum chromodynamics (QCD) through heavy-ion collider experiments, numerical simulations and theoretical model formalisms [90]. These FBs in the QCD circumstances are produced through heavy-ion collisional processes with energy on the GeV-scale [90], such as in the Relativistic Heavy-Ion Colliders (RHICs), etc. In contrast to the classical laboratory plasma FBs of size on the centimeter order, these FBs, produced in the QCD scenarios, are typically of the femtometer order [91]. It further adds to the applications of the FB research at subatomic scales against bulk macroscopic phenomena.

# 6 Comparative analysis

It is evident from the above discussion that the dynamics of FB formation and evolution in different terrains have similarities in different aspects of both pure and applied value. However, there still exist several characteristics distinguishing RFBs, IFBs, and associated sheath structures developed in one terrain to those found in another as presented in Tables-1-3. We highlight the main distinctions between RFB versus IFB (Table-1), laboratory plasma FB versus astrocosmic FB (Table-2), and regular sheath versus inverted sheath (Table-3), presented as follows.

**Table 1** Comparison of RFB and IFB

| S. No. | Item | RFB (Source) | IFB (Source) |
|---|---|---|---|
| 1 | FB location | On electrode (anode) surface [4, 6] | Inside electrode (anode) [8] |
| 2 | Electrode morphology | Solid (non-grid) [4, 6] | Reticular (grid) [8] |
| 3 | Glow region | Sheath region and Whole FB region [4, 6] | Within electrode (but not so in sheath) [8] |
| 4 | Collision and ionization region | Outside the electrode [4] | Inside the electrode (anode) [8] |
| 5 | Equipotentiality Electric potential | Radially non-existent [1] Radially dependent | Inside electrode [8] Constant inside the IFB |
| 6 | Electron density | Variable in entire FB [1, 4] | Gaussian inside and constant just outside electrode [8] |
| 7 | FB geometry | Independent of anode geometry [4] | Dependent on anode geometry [8] |
| 8 | Ripple factor | More in FB current [6] | Less in FB current [13] |
| 9 | Magnetic FB | Extensively studied [4] | Infancy stage [1] |
| 10 | Spatial dimension | Predominantly dependent on anode volume and pressure [4] | Determined by reticular anode grid [8] |
| 11 | Astro-natural relevancy | Strong [1, 3] | Weak [1] |

| 12 | Dominant instabilities | TSI, SPI, SII, PRI, RTI [1] | KHI, SPI, SII [1] |

**Table 2** Comparison of laboratory FB and astrocosmic FB

| S. No. | Item | Laboratory FB (Source) | Astrocosmic FB (Source) |
|---|---|---|---|
| 1 | Force causing glow | Electromagnetic [4] | Gravitothermal [2] |
| 2 | Interacting particles | Electrons and neutrals [4] | Neutral gas molecules [45] |
| 3 | Brightness | Less (compared to astrocosmic) | More (~ -3 in magnitude) [45] |
| 4 | Spatial size | 5-10 cm [4] | 1 $\mu$m -1 m [1] |
| 5 | Electrostatic sheath | Forms [4] | Does not form |
| 6 | DL | Forms [4] | Does not form |
| 7 | FB formation and location | Across and within anode [4,6] | Forms without any anode |
| 8 | Stability | Stable in given conditions [4] | Normally unstable [45] |
| 9 | Occurrence | Specific plasma conditions [4] | Natural [45] |
| 10 | Excitation mechanism | Collisional [4] | High-pressure [45] |
| 11 | Shape | Spherical or elongated (under magnetic field) [4, 6] | Always elongated (tailed) [45] |
| 12 | Plasma Eigenmodes | Electron and ion cyclotron waves [12, 13] | Yet to be known |
| 13 | Current carriers | Whistler and Alfven modes (electric) [28, 30] | Macroscopic neutral pressure (gaseous) [45] |
| 14 | Magnetic field | Influences geometric shape due to the presence of charges [1] | Inferior influence due to neutral atoms [45] |
| 15 | Symmetry | Spherical symmetric (with spherical anode) [4] | Axisymmetric [43, 45] |
| 16 | Efficiency | Less efficient due to less collisional cross-section [8] | More efficient due to greater collisional cross-section [45] |
| 17 | Collision nature | Inelastic due to electron energy absorbance by neutrals [4, 6] | Almost elastic due to mere gravitational pressure |
| 18 | Characteristic readings | Plasma probes (Langmuir, double, etc) [4, 8] | Astrospectrometric observations [45] |
| 19 | Wide utility | Plasma medicines, supernovae, gamma-ray bursts, etc. [1, 3, 92] | Yet to be known |
| 20 | Research activity | Extensive [1] | In infancy stage |
| 21 | Dominant instabilities | Transit-time instability, SPI, vircator-like instability, RTI, KHI, TSI, etc. [13, 14] | Streaming instability, RTI [70], Farley-Buneman [93-95] |

**Table 3** Comparison of regular sheath and inverted sheath

| S. No. | Item | Regular sheath | Inverted sheath |
|---|---|---|---|
| 1 | Net polarity | Positive [35] | Negative [43, 44] |
| 2 | Composition | Electrons, positive ions [45] | Electrons, negative ions [44] |
| 3 | Velocity distribution | Maxwellian [5] | Half-Maxwellian [43] |
| 4 | Formation criteria | Bohm criterion [35] | No Bohm-like criterion [43] |
| 5 | Electrode potential | Constant (homogeneous) [4] | Floating [43] |
| 6 | Charge emission | Dominantly constant [4] | Fluctuating due to pulses [43] |
| 7 | Heavy-ion sputtering | High [43] | Low [43] |
| 8 | Secondary ionization | Assumed [1] | Ignored [43, 44] |
| 9 | Applicability of Bohm criterion | Yes [35] | No [43, 44] |
| 10 | Formation time | Relatively longer [43] | Relatively shorter [43, 44] |
| 11 | Research activity | Extensive [1] | In infancy stage [43, 44] |

## 7 Conclusions

This meta-analytic review work focusses on a systematic overview to reveal the physical insights and interdisciplinary technological glimpses of existing PFS model analyses followed by closely associated diversified instability phenomena illustratively. The PFS investigation is apparently one of the most interesting but least addressed topics in the emerging arenas of plasma physics, engineering, and technology [1]. In a broader sense, it has stupendous applications in diversified domains ranging from nano-to-cosmic scales of space and time. The available investigations conducted on the PFS system and relevant plasma instabilities have reported several noticeable outcomes bearing widespread practical implications and applications [10]. As an instant example, the PFS system analyses reveal the formation of various linear and nonlinear waves, instabilities, and eigenmode structures generated with the help of a pulsing, DC, and AC central electrode in an external magnetic field [9]. It incorporates the electron-acoustic, ion-acoustic, Alfven, and whistler modes as the most prominent ones. The various PFS analyses emphasize one of the most crucial regions widely encountered in laboratory plasmas, i.e., plasma sheath and DL structures. In brief, the PFS analyses have led to a comprehensive understanding of collective waves, instabilities, their experimentations, diagnostic analyses, practical applications, and so on. The various required plasma parametric values, such as pressure, density, electric potential, and temperature, are also discussed extensively for respective applications in a broader horizon.

We further present tabular portrayals to highlight the main distinctions between RFB versus IFB (Table 1), laboratory plasma FB versus astrocosmic FB (Table 2), and regular sheath versus inverted sheath (Table 3) for a clear conceptualization.

Based on the above-mentioned scenarios, it can now be inferred that there still exists a wide area of such PFS-excited phenomena yet to be well investigated illustratively. A good number of experimental outcomes, previously reported in the literature [1], need further emboldened theoretical support. In this context, one can briefly mention a few observational aspects requiring requisite explanations from a wider theoretical viewpoint. As far as widely seen in the literature [2, 32, 49, 52, 86, 96], a few problems yet to be well investigated in this horizon may be summarily outlined as follows:

(a) A formalism development depicting the spatiotemporal velocity distribution of the plasma constituents (electrons and ions) near PFS systems [49];
(b) Mechanism of heat energy transfer from the ambient plasma to the constitutive neutrals in PFS systems [32];
(c) The PFS research has an extensive scope in interdisciplinary areas, such as plasma propulsion [87], plasma medicine [96-100], isotopic separation from a heterogenous gaseous mixture [86], quark and antiquark modelling [101-105], strongly nonlinear PFS instability [2, 52], etc.

As a future scope, the lack of substantial literature reporting on the application of pulsed electrode and magnetic field across the IFB motivates us in developing experimental data-based theories to understand the IFB behaviors in such special circumstances. Moreover, there exist insufficient reports in the literature on the wave and eigenmode formation in the case of IFB environments, unlike the RFBs, as in the literature. It is worth mentioning that we are planning to analyze the IFB system theoretically with the available experiential outcomes in future research [1, 8]. It would primarily aim at the excitation processes of possible collective oscillations or eigenmode formations extensively relevant in the ion-plasma frequency domains with both simpler bifluidic [54] and complex multifluidic (as pre-realized in other plasma systems) approaches [96]. Finally, we are herewith optimistic about the newer possibilities for further non-trivial FB exploration alongside the futuristic technological scope veiled in it.

**Acknowledgements** We appreciate the helps and support received from our laboratory fellow members, Astrophysical Plasma and Nonlinear Dynamics Research Laboratory (APNDRL), Department of Physics, Tezpur University, India. One of the authors, PKK, especially thanks the Inter-University Centre for Astronomy and Astrophysics (IUCAA), Pune, India, for academic associateship.

**Author contributions** All the authors have contributed significantly to the development of the current version of the proposed manuscript. Subham Dutta has performed the necessary calculations under the active supervision of Pralay Kumar Karmakar. Ahmed Atteya has contributed to the development of graphical, schematic, and visual analyses. The final refinement, proof-reading, and technical enrichment of the manuscript have been done jointly by all the authors.

**Funding** This research received no external funding.

**Data availability** No datasets were generated or analyzed during the current study.

## Declarations

**Ethics approval and consent to participate** Not applicable.

**Consent for publication** Not applicable.

**Competing Interests** The authors declare no competing interests.

**Correspondence** and requests for materials should be addressed to Pralay Kumar Karmakar.

**Clinical trial number** Not applicable.